\documentclass[aps,twocolumn,amsfonts,amsmath,amssymb,mathtools,cuted,prx,superscriptaddress,preprintnumbers,10pt,floatfix,reprint,nofootinbib]{revtex4-2}
\usepackage{graphicx,subfigure}
\usepackage{pictexwd,dcpic}
\usepackage{microtype}
\usepackage[hidelinks,colorlinks=true,linkcolor=blue,citecolor=blue]{hyperref}
\usepackage[utf8]{inputenc}
\usepackage{uniinput}
\allowdisplaybreaks

\usepackage{tikz-cd} 
\usetikzlibrary{backgrounds, arrows,calc,shapes,decorations.pathreplacing, decorations.markings, automata,positioning,matrix}

\newcommand{\be}{\begin{equation}}
\newcommand{\ee}{\end{equation}}
\newcommand{\ba}{\begin{aligned}}
\newcommand{\ea}{\end{aligned}}

\newcommand{\cE}{\mathcal{E}}
\newcommand{\cM}{\mathcal{M}}

\usepackage{ulem}
\makeatletter
\renewcommand{\p@subsection}{}
\renewcommand{\p@subsubsection}{}
\makeatother

\begin{document}

\title{SymTFTs for Continuous non-Abelian Symmetries}

\bigskip

\author{Federico Bonetti}
\affiliation{Department of Mathematical Sciences, Durham University}

\author{Michele Del Zotto}
\affiliation{Department of Mathematics,  Uppsala University,  SE-75120 Uppsala, Sweden}
\affiliation{Department of Physics and Astronomy,  Uppsala University,  SE-75120 Uppsala, Sweden}
\affiliation{Centre for Geometry and Physics,  Uppsala University,  SE-75120 Uppsala, Sweden}

\author{Ruben Minasian}
\affiliation{Institut de Physique Th\'eorique, Universit\'e Paris-Saclay, CNRS, CEA, F-9119, Gif-sur-Yvette,
France}
\begin{abstract}
Topological defects and operators give a 
 far-reaching
generalization of symmetries of quantum fields.  An auxiliary  topological field theory in one dimension higher than the QFT of interest, known as the SymTFT, provides a natural way for capturing such operators. This gives a new perspective on several applications of symmetries, but fails to capture continuous non-Abelian symmetries. The main aim of this work is to fill this gap.  Guided by geometric engineering and holography, we recover  various known features of representation theory of the non-Abelian symmetry from a SymTFT viewpoint. Central to our approach is a duality between the flat sector of Yang-Mills at zero coupling and non-Abelian BF theories. Our results extend naturally to models without supersymmetry.
\end{abstract}
\maketitle
\section{Introduction and motivation}
Symmetries provide essential underpinnings of our understanding of quantum systems.
Remarkably, 
in recent years the very notion of symmetries in quantum field theory (QFT) has undergone a revolution, based on the idea that symmetries are encoded in topological subsectors of the spectrum of QFT operators \cite{Gaiotto:2014kfa}.
 While ordinary symmetries are described by groups, symmetries characterized by topological operators have a higher structure that generalizes the notion of group multiplication, believed to be inscribed in a \textit{symmetry (higher) category} (see e.g.~\cite{Cordova:2022ruw,McGreevy:2022oyu,Gomes:2023ahz,Schafer-Nameki:2023jdn,Brennan:2023mmt,Bhardwaj:2023kri,Shao:2023gho,Carqueville:2023jhb} for recent reviews).
 
Building on the case of
\emph{finite}  generalized symmetries,
a compelling picture has emerged:
the symmetries of a
$d$-dimensional QFT    $\mathbf T$ are encoded
in an auxiliary topological
bulk-boundary system in $d+1$
 dimensions. This is known as
the \textit{topological symmetry theory} construction (or \textit{SymTFT} for short) \cite{Ji:2019jhk,Gaiotto:2020iye,Apruzzi:2021nmk,Freed:2022qnc} (see also \cite{Gukov:2020btk,DelZotto:2022ras,Bashmakov:2022jtl,Kaidi:2022cpf,Bashmakov:2022uek,Kaidi:2023maf,Chen:2023qnv,Bashmakov:2023kwo,Antinucci:2022cdi,Bhardwaj:2023ayw,Bartsch:2023wvv,Sun:2023xxv,Cordova:2023bja,Antinucci:2023ezl,Bhardwaj:2023fca,Bhardwaj:2023idu,Bhardwaj:2023bbf}). 
The SymTFT picture 
provides a powerful unifying
framework that encompasses
several crucial aspects
of the physics of symmetries, including: action on charged objects and structure of  multiplets; anomalies and 't Hooft anomaly matching; organization of gapped and gapless phases.
These considerations provide strong motivation to pursue the SymTFT program beyond finite symmetries. In this work, we
aim at extending this program to ordinary continuous non-Abelian symmetries, with the goal of
paving the way to exploring
non-finite generalized symmetries. This is a key result towards the understanding of the most general symmetry structures of QFTs.

The SymTFT construction is based on four ingredients: (1) a $d$-dimensional field theory $\widehat{\mathbf{T}}$ defined relative to (2) a bulk $d+1$ topological field theory $\mathbf{S}$, together with (3) a topological gapped $d$-dimensional boundary condition for $\mathbf{S}$, that we denote $\mathbf{B}$ such that there is (4) an isomorphism of field theories
\begin{equation}\label{eq:freedsandwitch}
\begin{gathered}
\scalebox{0.5}{
    \begin{tikzpicture}
        %\shadedraw [shading=axis,top color = red, bottom color=white, white] (0,0) rectangle (5,3);
        \draw[dotted] (0,0) -- (5,0); 
        \draw[dotted] (0,3) -- (5,3);
        \draw [very thick] (0,0) -- (0,3);
        \draw [very thick] (5,0) -- (5,3);
        \node[left] at (0,1.5) {{\Large $\mathbf{B}$}};
        \node[right] at (5,1.5+0.08){{\Large $\widehat{\mathbf{T}}$}};
        \node[above] at (2.5,1.2){{\Large $\mathbf{S}$}};
        \draw [very thick] (7.5,0) -- (7.5,3);
        \node[above] at (6.5,1.5){{\huge $\simeq$}};
        \node[right] at (7.5,1.5){{\Large $\mathbf{T}$}};
    \end{tikzpicture}}
    \end{gathered} \;\; . 
\end{equation}
The key point of this construction is that it 
provides a 
bulk-boundary realization of the
symmetry operators,
furnished by topological operators in $(\mathbf{S}, \mathbf{B})$
and independent on the specific
quantum system 
$\widehat{\mathbf T}$
on which the symmetry acts.
Dually, $\mathbf B$ captures background configurations for the symmetry of $\mathbf T$, which is gauged in the bulk (where only the Drinfeld center of the symmetry is manifest).

In this letter, our main goal
is to provide a bulk-boundary
realization of symmetry operators
for continuous global
symmetries. Recent proposals in the literature address this topic \cite{Brennan:2024fgj,Antinucci:2024zjp} advancing a SymTFT for continuous $U(1)$ symmetries and conjecturing a possible generalization for continuous non-Abelian symmetries in terms of a non-Abelian version of BF theory. From this latter perspective, however, it is unclear how the representation theory of the non-Abelian flavor symmetry is realized as a bulk-boundary system.

In this letter we give a dual formulation of the topological symmetry theory in terms of the flat sector of non-Abelian Yang-Mills theory in the limit of zero gauge coupling, that allows to address this open question among others.

Let us emphasize an important difference between the SymTFT setup for finite symmetries, and the setup of this work.
In the finite case,
the SymTFT furnishes also
a systematic characterization of 
possible gaugings of a global symmetry. Indeed,
gauging a finite symmetry is a topological operation encoded in a topological boundary-changing interface
of $\mathbf{S}$, keeping 
$\widehat{\mathbf{T}}$
fixed.
In contrast, gauging a continuous symmetry 
introduces new degrees of freedom in the spectrum (the gauge mediators as well as various topological solitons)
and cannot be captured by topological manipulations only.
Nevertheless, it is still
fruitful to pursue a bulk-boundary viewpoint on continuous symmetries. 
(We comment further on gauging in Section \ref{sec:topop}.)

Our proposal is realized explicitly in the context of geometric engineering as well as in holography, that serve as examples. We stress that the structure which is responsible for capturing the symmetry is the LHS of the diagram in Equation \eqref{eq:freedsandwitch}, namely the data (2) and (3) in the SymTFT construction, that are completely independent from the choices of $\mathbf{T}$ as well as $\widehat{\mathbf{T}}$:
\begin{equation}
\begin{gathered}
\scalebox{0.5}{
    \begin{tikzpicture}
        \draw[dotted] (0,0) -- (5,0); 
       \draw[dotted] (0,3) -- (5,3);
      \draw [very thick] (0,0) -- (0,3);
       \node[left] at (0,1.5) {{\Large $\mathbf{B}$}};
        \node[above] at (2.5,1.2){{\Large $\mathbf{S}$}};
    \end{tikzpicture}}
    \end{gathered} \;\; .
\end{equation}
Here we exploit the extra features (such as superconformal invariance) as a strategy for obtaining an explicit expression for $\mathbf{S}$. By the above argument, these extras are not strictly necessary, and indeed our construction holds for non-supersymmetric theories as well.\footnote{Similar to the Abelian case discussed in \cite{Brennan:2024fgj,Antinucci:2024zjp}, our analysis applies to   \textit{flat} backgrounds of the continuous symmetry. These backgrounds are captured by the boundary conditions of our bulk TFT. We reserve a better treatment including more general backgrounds for future work.}

This note is organized as follows. In Section \ref{sec:duality} we discuss the relation between the Horowitz non-Abelian BF theory \cite{Horowitz:1989ng} and 
the flat sector of Yang-Mills theory in the limit of zero coupling. In Section \ref{sec:evidence} we present evidence for our proposal. Our main classes of examples are SCFTs realized via geometric engineering in M-theory, discussed in \S~\ref{sec:geomengo}, as well as holographic SCFTs, discussed in \S~\ref{sec:holography}. In particular, our argument applies uniformly to all holographic SCFTs of $d\geq 4$, as expected since $d+1=4$ is a critical dimension for gravitational theories in AdS. Our analysis of free Yang-Mills as a SymTFT for continuous non-Abelian symmetries is complemented by a first series of applications discussed in Section \ref{sec:topop}, where we discuss how to recover the representation theory of non-Abelian Lie groups from the bulk-boundary perspective. We present some conclusions and directions of further study in Section \ref{sec:conclusions}.

\medskip

\noindent
\textbf{Note Added.} We were informed by Fabio Apruzzi of an upcoming work \cite{Apruzzi:2024htg} that has some overlap with the results in our note.

\section{SymTFT for continuous non-Abelian symmetries}\label{sec:duality}
Motivated by geometric engineering in string theory and holography as discussed in Section \ref{sec:evidence} below, we propose that the SymTFT for a QFT in $d$-dimensions with a continuous non-Abelian symmetry of type $G$ contains a $(d+1)$-dimensional free Yang-Mills theory, namely a Yang-Mills theory at zero gauge coupling $g \to 0$. Our claim might seem in tension
with the recent proposal by \cite{Brennan:2024fgj,Antinucci:2024zjp} that the relevant SymTFT is the non-Abelian Horowitz BF theory. In this Section we briefly outline a duality of free flat Yang-Mills with the Horowitz non-Abelian BF theory thus reconciling our proposals. 

Following Witten (see e.g.~\cite{Witten:1992xu}, sec.~4.2), one can re-write the Yang-Mills action as follows:
\begin{align} \label{eq_original}
&\mathbf{S} \supset  - {1\over 2 g^2} \int_{M_{d+1}} \text{tr} \left(f_{2} \wedge \ast f_{2}\right)\\
&= \int_{M_{d+1}} \bigg[ \text{tr} \left(f_{2} \wedge h_{d-1}\right) - \frac {g^2}{2} \text{tr} \left( h_{d-1}  \wedge \ast h_{d-1}\right) 
\bigg] \ ,
\nonumber 
\end{align}
where $h_{d-1}$ is a collection of Lagrange multiplier $(d-1)$-form fields transforming in the adjoint of $\mathfrak{g}$. 

The corresponding equation of motion is valued in the adjoint of $\mathfrak{g}$ and gives $f_{2} = g^{2} \ast  h_{d-1}$. Motivated by geometric engineering and holography, to obtain the contribution to the symmetry theory we need to take  the limit $g^2 \to 0$ of the action above, in particular this implies only the flat connections contribute to the symmetry theory.

By the Ambrose-Singer theorem, all the holonomies of such gauge fields are topological operators, hence we expect to have topological operators corresponding to the Wilson lines of $\mathfrak g$ Yang-Mills. 

In the $g \to 0$ limit, we obtain the action
\begin{equation} \label{eq_nonAb_BF}
 \int_{M_{d+1}} \text{tr} \left(f_{2} \wedge h_{d-1} \right) \ , 
\end{equation}
which is exactly the non-Abelian Horowitz BF theory \cite{Horowitz:1989ng}.
In the original action
\eqref{eq_original}, the $(d-1)$-form
$h_{d-1}$ has no gauge redundancy,
but in \eqref{eq_nonAb_BF}
$h_{d-1}$ acquires a gauge redundancy,
\be \label{eq_gauge}
h_{d-1} \sim h_{d-1} + D\lambda_{d-2} \ , 
\ee 
where $D$ is the exterior covariant derivative in the adjoint representation.
The topological action 
\eqref{eq_nonAb_BF} is invariant under
\eqref{eq_gauge} by virtue
of the Bianchi identity
$Df_2=0$. We stress that the $g^2\to 0$ limit of Yang-Mills is a free theory of gapless gluons. The topological BF action \eqref{eq_nonAb_BF} captures a sector of this theory with only flat connections, where the propagating gluons are indeed absent.

As already pointed out in \cite{Brennan:2024fgj,Antinucci:2024zjp}, from the perspective of non-Abelian BF theory one would like to define holonomies for the Wilson surfaces of $h_{d-1}$, but there is not a good notion of such ``surface ordered'' integrals. Here our duality with the flat sector of YM at zero coupling suggests a way out. We know YM has Gukov-Witten (GW) operators, which are codimension 2 in the bulk symmetry theory, and become topological as well in the limit of zero coupling. This prompts us to look for analogous disorder operators in the Horowitz BF theory. In YM at zero-coupling one can use the description of the GW operators in terms of the dual fluxes for the Cartan subalgebra of the gauge group modulo Weyl \cite{Cordova:2022rer,Antinucci:2022eat}. In particular, from this perspective, we expect the GW operators in the bulk give rise to a non-invertible subalgebra of the spectrum of the SymTFT for non-Abelian symmetries.

\section{Evidence from geometric engineering and holography}\label{sec:evidence}
In this section we discuss evidence for our claim arising from geometric engineering examples (\S~\ref{sec:geomengo}), as well as holographic SCFTs (\S~\ref{sec:holography}).

\subsection{SymTFT from Geometry}\label{sec:geomengo}

For the sake of brevity, in this letter we consider the simplest possible geometric engineering scenario, where a $d$-dimensional SCFT is obtained by geometric engineering M-theory on a conical Calabi-Yau singularity $X$ with metric
\begin{equation}\label{eq:conical}
ds^2_X = dr^2 + r^2 ds^2_{L_X} \ , 
\end{equation}
where $L_X$ is a Sasaki-Einstein manifold, also known as the link of the singularity. We use this class of geometries here as a main source of examples, but we stress that our results are more general.\footnote{\textit{Mutatis mutandis} our results generalize straightforwardly to type IIA geometric engineering limits, and little more effort is needed in the context of IIB or Heterotic scenarios, which we reserve to study in future work.} As a example consider the case $X$ is an orbifold of $\mathbb C^n$, $\mathbb C^n /\Gamma$, where $\Gamma$ is a finite subgroup of $SU(n)$ preserving the Calabi-Yau structure. In these cases, the link is a lens space which has at worse orbifold singularities $L_X = \mathbb S^{2n-1}/\Gamma$. Many examples of $d$-dimensional SCFTs are obtained in this way, in particular choosing $n=3$, one obtains the 5d orbifold SCFTs from M-theory \cite{Xie:2017pfl, Acharya:2021jsp,Tian:2021cif,DelZotto:2022fnw}.

\medskip

The geometric origin of the SymTFT for finite symmetries in these cases is encoded in a ``geometric engineering limit at infinity'', where the topological operators of $\mathbf{S}$ have a geometric origin as membranes wrapped on torsional cycles of $L_X$ 
\cite{Apruzzi:2021nmk,vanBeest:2022fss, Apruzzi:2022rei, GarciaEtxebarria:2022vzq, Heckman:2022muc, Heckman:2022xgu,Etheredge:2023ler, Dierigl:2023jdp, Cvetic:2023plv, Lawrie:2023tdz,Bah:2023ymy, Apruzzi:2023uma,Baume:2023kkf,Yu:2023nyn,Heckman:2024oot,  Heckman:2024obe, DelZotto:2024tae}
building on \cite{DelZotto:2015isa,GarciaEtxebarria:2019caf,Morrison:2020ool,Albertini:2020mdx,Hubner:2022kxr}.
The case of continuous symmetries deserve a special analysis. In many cases, continuous symmetries are realized in geometry as non-compact curves of singularities, that intersect the singular locus where the SCFT is located (see e.g.~\cite{Benini:2009gi,DelZotto:2014hpa,Hayashi:2019fsa,Eckhard:2020jyr,Bhardwaj:2020ruf,Bhardwaj:2020avz,Apruzzi:2021vcu,DelZotto:2021ydd,DelZotto:2022joo,Acharya:2023bth,DeMarco:2023irn} 
for some examples in various spacetime dimensions). More precisely, consider the case of continuous non-Abelian 0-form symmetries, that arise from non-compact loci of du-Val singularities (split or non-split corresponding to simply-laced vs non-simply laced groups). Since these non-compact loci extend at infinity, the resulting link $L_X$ is also singular, with \textit{compact} loci $C_{\mathfrak{g}_i} \subset L_X$ of du-Val singularities. The latter correspond to \textit{gauge} theories in one higher dimension, with gauge couplings $1/g^2_{i} \sim \text{vol } (C_{\mathfrak{g}_i})$.
However, since the link is at infinity in geometry, all of its compact cycles have infinite volume in the metric \eqref{eq:conical}, hence the corresponding higher dimensional gauge theory contributes to the SymTFT only in the limit $g^2_{i} \to 0$. To capture the contribution to the symmetry theory from the singularity link we can follow the procedure outlined in the previous section for each of these factors, obtaining the following action,\footnote{It would be interesting to understand whether the emergent gauge symmetry in \eqref{eq_gauge} has a higher dimensional origin. A possible approach could be provided by the symmetry descent \cite{GarciaEtxebarria:2024fuk,Gagliano:2024off}. However in geometric engineering it is typically very hard to detect non-Abelian structures---one proceeds indirectly, e.g. engineering a Coulomb phase where the non-Abelian W-bosons originate from the low energy limit of wrapped membranes. We expect similar subtleties to arise here.}
\begin{equation} \label{eq_nonAb_BF2}
 \int_{M_{d+1}} \text{tr} \left(f_{2,i} \wedge h_{d-1,i} \right) \ . 
\end{equation}

\medskip

\noindent\textbf{An example: 5d $T_N$ SCFTs.}  The 5d $T_N$ theory is obtained geometrically engineering M-theory on the orbifold $X_N = \mathbb C^3 / (\mathbb Z_N \times \mathbb Z_N)$ where the action by $\mathbb Z_N \times \mathbb Z_N$ is $(z_1,z_2,z_3) \to (\alpha z_1, \overline{\alpha}  \beta z_2,\overline{\beta}z_3)$ and $\alpha^N = \beta^N=1$ \cite{Benini:2009gi}.
In this case the three curves $C_1 \colon z_2 = z_3 = 0$, $C_2 \colon z_1 = z_3 = 0$ and $C_3 \colon z_1 = z_2 = 0$ correspond to three different non-compact loci of $\mathbb C^2/\mathbb Z_N$ singularities, associated to the 0-form symmetry $SU(N)^3$ of this theory. Consider the link of this singularity, one obtains $
L_X = \mathbb S^5 / (\mathbb Z_N \times \mathbb Z_N )\,,
$ where the $\mathbb S^5$ has equation $|z_1|^2 + |z_2|^2 + |z_3|^2 = R^2 $ and the limit $R\to \infty$ is taken in the metric. 

Each of the three curves of singularities in $X_N$, corresponds to a singular locus on $L_X$, $C_i \, \longleftrightarrow \, z_i \bar z_i = R^2$, which is an $S^1$ with radius $R$.

The reduction of M-theory on $S^1 \times \mathbb C^2/\mathbb Z_N$ correspond to a 6d \textit{dynamical} $\mathfrak{su}_N$ field theory with gauge coupling ${1 \over g^2} \sim R$. In the spirit of the geometric engineering at infinity origin of the SymTFT, it is tempting to conclude that each of these fixed loci contributes one such gauge theory sector. However, one has to keep track of the fact that the link of the singularity in the metric \eqref{eq:conical} is of infinite size.\footnote{All local degrees of freedom are expected to decouple in this limit, and one is left with a topological sector \cite{DelZotto:2024tae}.} In particular, we need to take the limit $R \to \infty$ in the analysis above, which is a limit of zero gauge coupling. To capture the contribution to the symmetry theory from the singularity link, then by the steps outlined above we obtain $\sum_{i=1}^3 \int_{M_6} \text{tr} \left(f_{2,i} \wedge h_{4,i} \right)$, as expected.

\subsection{SymTFT from Holography}\label{sec:holography}

In this section we confirm our main result for those CFTs in $d\geq 4$ dimensions with a continuous non-Abelian 0-form symmetry $G$ that have a holographic dual in $AdS_{d+1}$.\footnote{In the case $d=2,3$ the holographic dictionary is slightly more complicated, as emphasized by Harlow and Ooguri \cite{Harlow:2018tng}. In particular, for $d=2$ the analysis is subtler due to logarithmic scaling \cite{Faulkner:2012gt} and the possibility of quadratic Chern-Simons terms.} In AdS/CFT the usual interpretation is that the holographic dual of a global symmetry is indeed a gauge symmetry in the bulk \cite{Witten:1998qj}. In particular, the $AdS_{d+1}$ low-energy action contains a Yang-Mills term
\be 
S_{\rm YM} = - \frac{1}{2g^2} \int_{AdS_{d+1}} {\rm tr}(f_2 \wedge * f_2) \ , 
\ee 
as well as other interactions, depending on the model. To capture the holographic dual of the topological symmetry theory we consider an asymptotic limit of the bulk gravitational theory, focusing on the region close to the AdS boundary where only a topological sector of the bulk theory survives.\footnote{See e.g.
\cite{Bah:2019rgq,Bergman:2020ifi,Bah:2020uev,Bergman:2022otk,Apruzzi:2022rei,GarciaEtxebarria:2022vzq,Heckman:2022xgu,Etheredge:2023ler,  Bah:2023ymy, Apruzzi:2023uma,Heckman:2024oot,Heckman:2024obe},
building on
\cite{Witten:1998wy,Belov:2004ht}.}
We want to estimate the effect of the Yang-Mills term near the conformal boundary of $AdS_{d+1}$. We work in a fixed
pure $AdS_{d+1}$ background. The metric is conformally flat,
\be 
ds^2_{AdS_{d+1}} = \frac{L^2}{z^2} d\widetilde {s}^2_{d+1} \ , \quad 
d\widetilde {s}^2_{d+1} = dz^2
+ ds^2_{\mathbb R^{1,d-1}} \ ,
\ee 
with $L$ the $AdS_{d+1}$ radius
and the conformal boundary located at $z=0$.
We can write the Yang-Mills term as
\begin{align} \label{eq_AdS_action}
&S_{\rm YM} = \int_{AdS_{d+1}}
- \frac{1}{2g^2}  \frac {L^{d-3}} {z^{d-3}}  {\rm tr}(f_2 \wedge \widetilde * f_2) \\
&= \int_{AdS_{d+1}} \bigg[ 
{\rm tr}(h_{d-1} \wedge f_2)
- \frac{g^2}{2} \frac{z^{d-3}}{L^{d-3}}
{\rm tr}(h_{d-1} \wedge \widetilde * h_{d-1}) 
\bigg] 
\ , \nonumber
\end{align}
where $\widetilde *$ denotes
the Hodge star of the flat metric
$d\widetilde s^2_{d+1}$,
and we have introduced
the auxiliary field
$h_{d-1}$, a $(d-1)$-form 
with values in the  Lie algebra
of $G$.
As we approach the conformal
boundary $z \rightarrow 0$,
the ${\rm tr}(h_{d-1} \wedge \widetilde * h_{d-1})$ term is subleading.
An alternative, more physical viewpoint is the
following. We have a gauge theory in $AdS_{d+1}$.
Approaching the conformal boundary probes its low-energy regime. But a non-Abelian gauge theory in
dimension $\ge 5$ is IR free,
with the effective $g^2$ coupling going to zero as
$z \rightarrow 0$. If we neglect 
the ${\rm tr}(h_{d-1} \wedge \widetilde * h_{d-1})$
in \eqref{eq_AdS_action},
we recover the non-Abelian BF theory 
\be  \label{eq_ads_bf}
S_{\rm BF} = \int_{AdS_{d+1}}{\rm tr}(h_{d-1} \wedge f_2) \ . 
\ee 

\medskip

\noindent{\textbf{An example: the 6d (1,0) E-string theory.}} The E-string theory of rank $N$ is the 6d (1,0) SCFT of a stack of $N$ M5 branes probing one M9 brane in M-theory. The M9 brane has an $E_8$ gauge symmetry and the E-string inherits an $E_8$ global symmetry. The holographic dual of the E-string theory is given by $AdS_7 \times \mathbb S^4/\mathbb Z_2$ where $\mathbb Z_2$ acts by $(x_1,x_2,x_3,x_4,x_5) \mapsto(x_1,x_2,x_3,x_4,-x_5)$ \cite{Berkooz:1998bx}. There is a fixed $\mathbb S^3$ at $x_5=0$.  On such a boundary, by the Hořava-Witten construction, there is an M9 brane that by inflow supports a 10d $E_8$ gauge theory. The latter is reduced on $\mathbb S^3 \times AdS_7$. By the reasoning in this paper, the resulting $E_8$ gauge fields in $AdS_7$ have a SymTFT action that contains the term \eqref{eq_ads_bf} corresponding to the $E_8$ global symmetry of the 6d (1,0) rank $N$ E-string SCFT (at large $N$).

\medskip

\noindent{\textbf{Remarks:}}
\begin{enumerate}
\item The charged operators in the CFT$_d$ that correspond to the end-points for the topological Wilson lines and Gukov-Witten operators of the $g^2=0$ gauge theory map to extended operators that end on the boundary and extend in the bulk. Away from the boundary the gauge coupling is not suppressed anymore and the extended gauge-theory operators are dressed by non-topological terms. For consistency with the completeness hypothesis in quantum gravity we expect all these operators end on charged objects in the bulk \cite{Polchinski:2003bq,Banks:2010zn}. These endings in the bulk are the holographic dual to the extended charged operators of the CFT. The boundary conditions at infinity are the holographic counterpart of the boundary conditions $\mathbf{B}$.
\item A more refined approach to symmetries indicates that \textit{splittable} global symmetries of the boundary CFT (which are the kind we are considering here due to the presence of Noether currents) are dual to a \textit{long-range} gauge symmetries in the bulk \cite{Harlow:2018tng}. This is consistent with our proposal: in the zero-coupling limit, the gauge mediators in the bulk are essentially photon fields valued in the Cartan of the gauge group (modulo Weyl), which are definitely long-range fields and indeed are the only ones that survive in the asymptotic region close to the AdS boundary.
\item From the holographic perpsective Abelian symmetries can be treated similarly, hence our analysis also confirms that the SymTFT for  Abelian $U(1)$ symmetries is indeed the one proposed in \cite{Brennan:2024fgj,Antinucci:2024zjp}.
\end{enumerate}

\section{Continuous non-Abelian topological symmetries}\label{sec:topop}

From the geometric engineering analysis above as well as from the holographic approach it is natural to interpret the SymTFT for continuous gauge symmetries of a $d$-dimensional QFT as a gauge theory in $d+1$ dimensions with gauge group $G$ at $g_{\rm YM} = 0$.\footnote{We are especially indebted to Kantaro Ohmori for a discussion that helped clarifying and shaping some key parts of this section, in particular,  how the non-Abelian symmetry is realized along the boundary.} The $(d+1)$-dimensional theory has the following topological extended operators:
\begin{itemize}
\item  Wilson line operators $W_1^{(\mathbf R)}$
labeled by an irreducible representation 
$\mathbf R$ of $G$;
\item $(d-1)$-dimensional Gukov-Witten \cite{Gukov:2006jk,Gukov:2008sn} operators
$T_{d-1}^{([g])}$
labeled by a conjugacy class $[g]$
of elements in $G$.
\end{itemize}
In the limit $g_{\rm YM} \rightarrow 0$,
all Gukov-Witten operators of $G$ Yang-Mills
theory become topological \cite{Antinucci:2022eat}.
We have argued above that
Yang-Mills theory in the 
limit $g_{\rm YM} \rightarrow 0$
is captured by the non-Abelian
BF theory \eqref{eq_nonAb_BF}.  
Then, $a_1$ is flat
and the Wilson lines become topological
as well.

Here, by Gukov-Witten operators in a $G$
Yang-Mills theory we mean the following: inserting
$T_{d-1}^{([g])}$ on $M_{d-1}$ 
corresponds to performing the path integral
on $G$ gauge field $a_1$ configurations
with a prescribed 
holonomy  
along the $\mathbb S^1$ linking $M_{d-1}$.
The holonomy is only specified up to conjugation.
This is necessary for bulk gauge invariance,
because the holonomy ${\rm hol}_\gamma(a_1)$ of $a_1$ along a closed
path $\gamma$ based at a point $P$
transforms as
${\rm hol}_\gamma(a_1) \mapsto h(P)^{-1}{\rm hol}_\gamma(a_1) h(P)$
under the gauge transformation
$a_1 \mapsto h^{-1}(A+d)h$.\footnote{It would be interesting to investigate the connection between 
the Gukov-Witten operators we discuss in this note and the non-invertible averages introduced by Córdova, Ohmori and Rudelius in \cite{Cordova:2022rer}.} 
\footnote{Gukov-Witten operators
for non-Abelian symmetries in the context of the holographic 
SymTFT have been recently discussed in~\cite{Heckman:2024oot}.}

We can adopt the same strategy to define
Gukov-Witten operators in the non-Abelian
BF theory \eqref{eq_nonAb_BF}.
If we adopt this viewpoint,
we avoid difficulties related to defining 
``surface ordered'' integrals of
$h_{d-1}$ \cite{Brennan:2024fgj}. Building on \cite{Cattaneo:1996pz,Cattaneo:2000mc,Cattaneo:2002tk}, in follow up work we reconcile these two constructions \cite{Bonetti:2025dvm}.

Having described the bulk theory $\mathbf S$
as the $g_{\rm YM} \rightarrow 0$ limit of $G$
Yang-Mills theory, or equivalently the
non-Abelian BF theory \eqref{eq_nonAb_BF},
let us now turn to the symmetry boundary
$\mathbf B$.  We define it by 
choosing Dirichlet boundary conditions
for the $G$-gauge field $a_1$.
Standard arguments then imply that
$\mathbf B$ carries a global $G$ symmetry.
In particular, 
we have topological $(d-1)$-dimensional
operators $S^{(g)}_{d-1}$
living on $\mathbf B$ and labeled by
a group element $g \in G$.

\begin{figure}
\centering
\includegraphics[width=3.6cm]{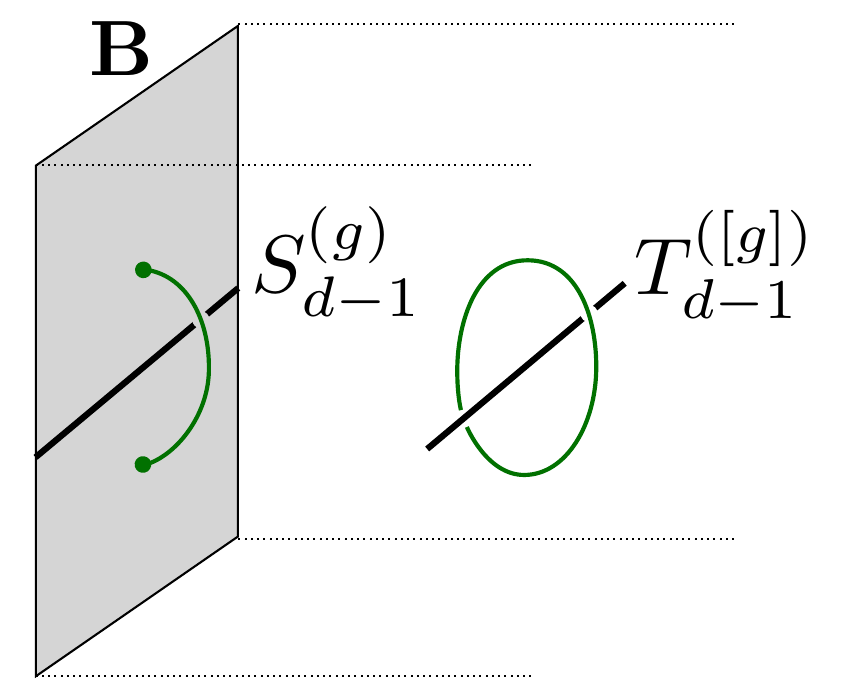}
\caption{A Gukov-Witten operator
$T_{d-1}^{([g])}$ in the bulk
induces a prescribed holonomy
along a small circle linking it.
The holonomy is $g$ up to conjugation.
The topological operator
$S_{d-1}^{(g)}$ is ``anchored''
to $\mathbf B$. It induces a prescribed
value $g$ for the quantity
$\mathrm{Pexp} \int_{P}^Q a_1$
where $P$, $Q$ are the endpoints of a small semicircle linking $S_{d-1}^{(g)}$. \label{fig_GW}}
\end{figure}

Crucially, in the bulk theory the Gauss's law constraint holds and, because of gauge invariance, we can only write down topological operators labeled by conjugacy classes in the bulk gauge group, that are gauge invariant. On the contrary, gauge transformation asymptote to the identity at the boundary, and for this reason a global 0-form $G$ gauge symmetry is realized.\footnote{This is reminiscent of the construction of the Drinfeld Center of a symmetry category.}

We can describe the topological
operators $S^{(g)}_{d-1}$ on $\mathbf B$
as  operators
of $(d+1)$-dimensional $G$ Yang-Mills
theory defined on a half-space.
The support $M_{d-1}$
of $S^{(g)}_{d-1}$
is linked by a semicircle.
Inserting $S^{(g)}_{d-1}$
means performing the path integral
over $a_1$ on the half-space,
with Dirichlet boundary conditions on
the boundary $\mathbf B$,
and imposing that the 
path-ordered exponential of the integral of $a_1$ along the semicircle
equals $g$. See Figure \ref{fig_GW}.
Crucially, this integral of $a_1$
on the semicircle is gauge invariant.
This is because we can only perform
gauge transformations in the half-space
that are trivial at the boundary $\mathbf B$,
and the endpoints of the semicircle
lie on $\mathbf B$.

We emphasize that the topological
operators $S^{(g)}_{d-1}$
are ``anchored'' to $\mathbf B$,
and cannot be brought into the bulk of
$\mathbf S$ (this would violate bulk
gauge invariance, per the discussion above).
We can, however, start with a 
Gukov-Witten operator $T_{d-1}^{([g])}$ 
in the bulk, and project it parallel
onto the boundary $\mathbf B$.
The resulting topological operator
on $\mathbf B$ is not simple,
but rather a sum of operators
$S^{(g')}_{d-1}$ with $g'\in [g]$.\footnote{This would be an averaged operator over the Haar measure of the group, reminiscent of the discussion in \cite{Cordova:2022rer}.} Only these averaged operators can move from the boundary into the bulk and viceversa, as genuine operators.\footnote{This is consistent with the fact that the operators $S^{(g)}_{d-1}$ would be codimension 2 in the bulk and if genuine, they could not have a non-Abelian fusion.} However, the operators $S^{(g)}_{d-1}$ can be moved into the bulk as non-genuine operators attached to a codimension 1 topological defect, similarly to what observed for topological lines with non-trivial braiding in \cite{Argurio:2024oym}. It would be interesting to use the methods developed in \cite{Bonetti:2025dvm} to study this further.

\begin{figure}
\begin{center}
\includegraphics[width=5.2cm]{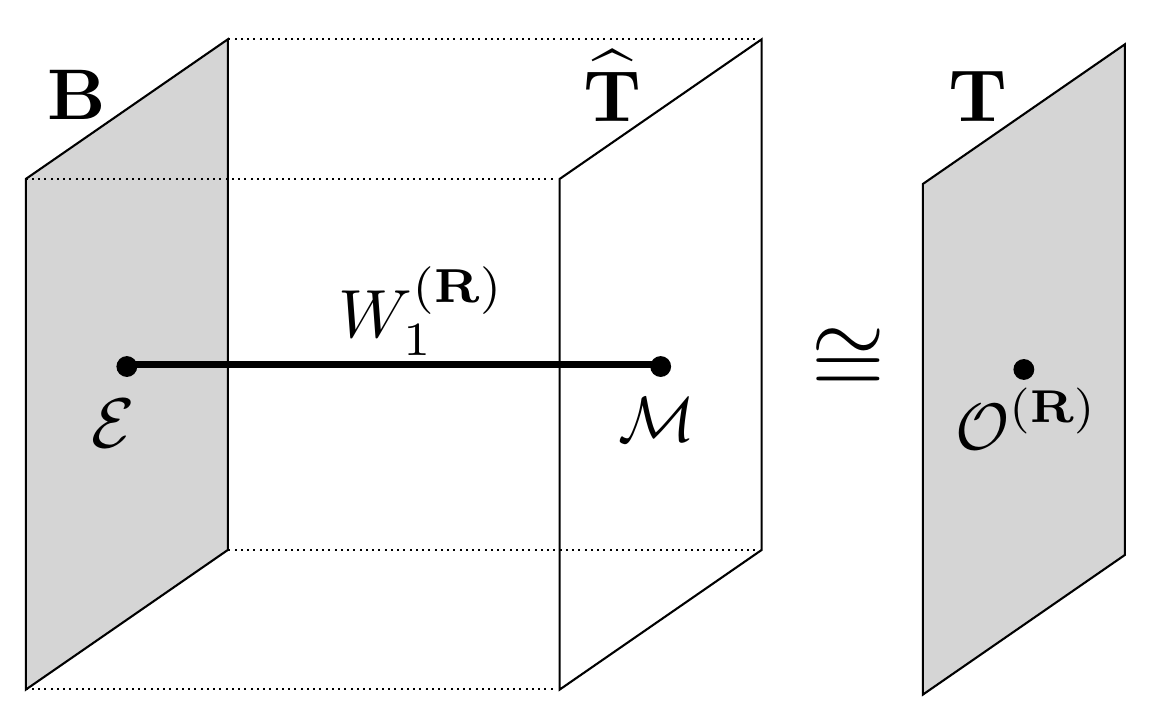}
\end{center}

\caption{Sandwich construction
for a genuine local operator
$\mathcal O^{(\mathbf R)}$ of the theory $\mathbf T$ in the representation
$\mathbf R$ in terms of a Wilson
line $W_1^{(\mathbf R)}$
stretched between
a topological endpoint $\mathcal E$
on $\mathbf B$ and a non-necessarily topological endpoint $\mathcal M$
on $\widehat{\mathbf {T}}$.
\label{fig_stretched}}
\end{figure}

Next, let us discuss charged local operators and how $G$
acts on them.
Generalized charges have been studied extensively in recent works 
\cite{Lin:2022dhv,Bartsch:2023pzl,Bartsch:2023wvv,Bhardwaj:2023wzd,Bhardwaj:2023ayw}.
In the remainder of this section
we follow closely the discussion in \cite{Lin:2022dhv,Bhardwaj:2023ayw}.
For simplicity, we restrict
to genuine local operators transforming in a representation $\mathbf{R}$.
In the SymTFT picture these operators are engineered
by stretching a Wilson line $W_1^{(\mathbf R)}$
along the interval direction,
connecting a point $\cM$ on $\widehat{\mathbf T}$
to a point $\cE$ on $\mathbf B$,
see Figure \ref{fig_stretched}.
This configuration is allowed
by our choice of Dirichlet boundary
conditions for $a_1$ on $\mathbf B$.

As stressed above,
the topological operators
$S_{d-1}^{(g)}$ labeled by elements of $G$
live on $\mathbf B$. Thus, the $G$-action
is encoded on the action of
$S_{d-1}^{(g)}$ on the endpoint $\cE$
of $W_1^{(\mathbf R)}$ on $\mathbf B$.
This is to be contrasted, for instance,
with the case of an Abelian   global
symmetry, for which 
the $G$-action can also be seen via linking
in the bulk.

\begin{figure}
\centering
\includegraphics[width=8.3cm]{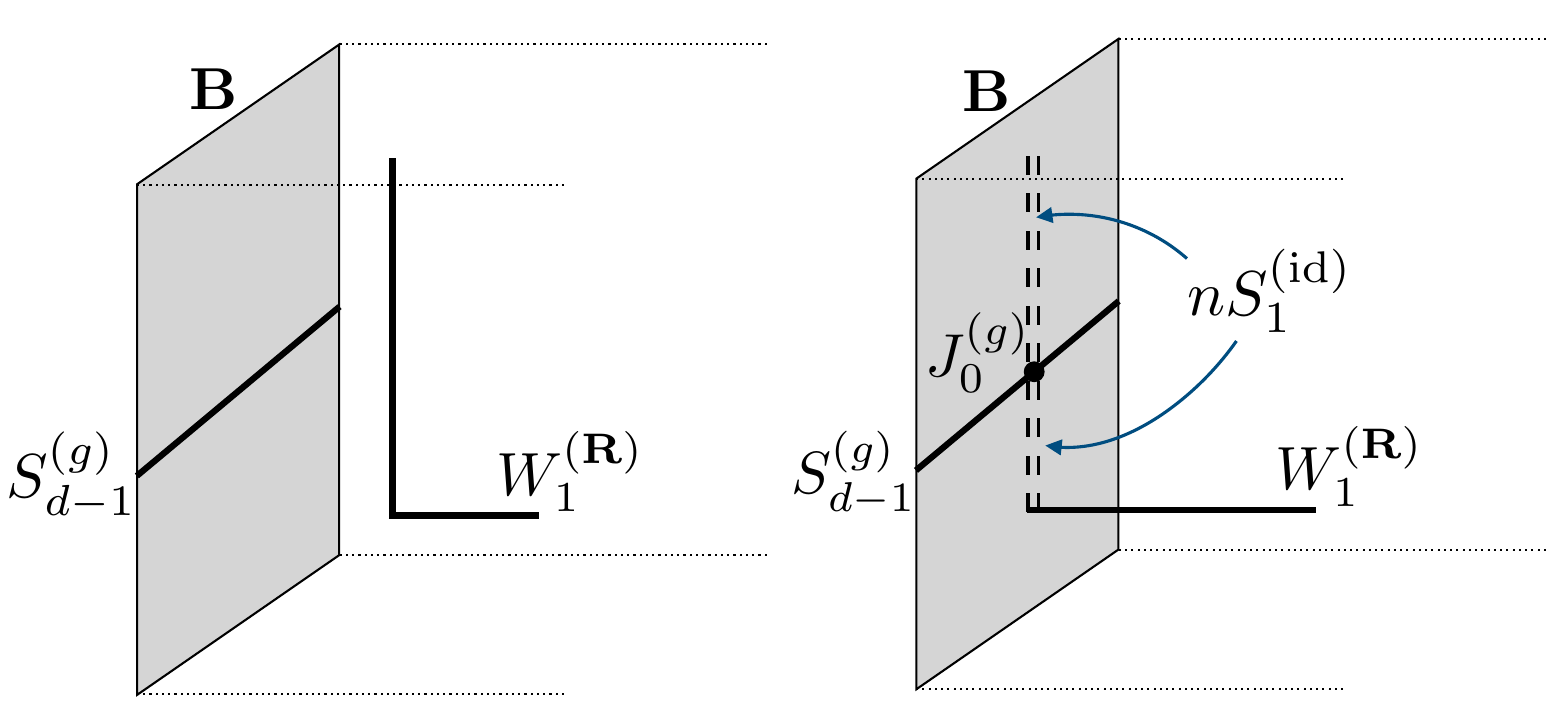}
\caption{When the Wilson line $W_{1}^{(\mathbf R)}$ is projected
parallel to the boundary $\mathbf B$
it yields $n S_1^{(\mathrm{id})}$,
denoted with a double dashed line.
Projecting an L-shaped configuration
for $W_{1}^{(\mathbf R)}$
gives the canonical endpoint
on $\mathbf B$ connecting
$W_1^{(\mathbf R)}$ and $n S_1^{(\mathrm{id})}$.
When $W_{1}^{(\mathbf R)}$ is projected 
on top of an operator $S_{d-1}^{(g)}$,
we get the point-like topological
junction $J_0^{(g)}$.
\label{fig_proj}}
\end{figure}

\begin{figure}
\centering

\begin{center}
\includegraphics[width=6.5cm]{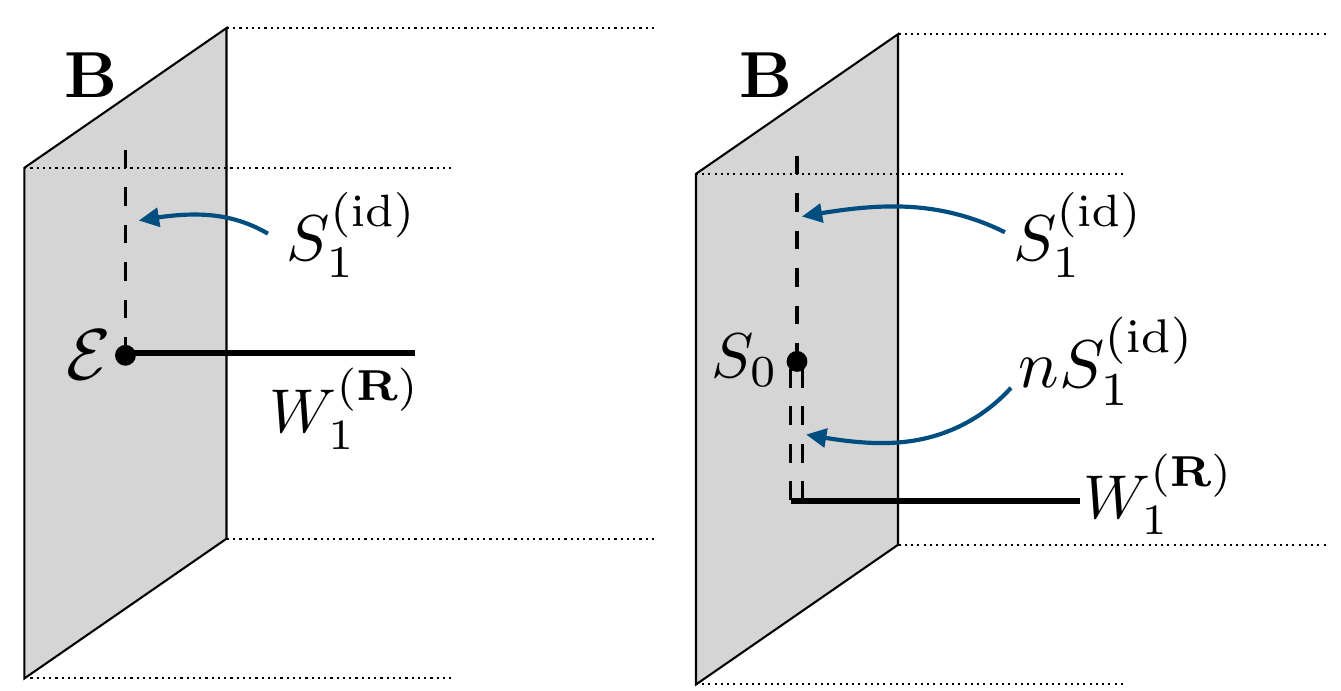}
\end{center}

\caption{The endpoint 
$\mathcal E$ of $W_1^{(\mathbf R)}$
is a junction with the trivial
line $S_1^{\rm (id)}$ on $\mathbf B$.
By projecting a segment of
$W_1^{(\mathbf R)}$
below $\mathcal E$,
we resolve $\mathcal E$
as a topological junction
$S_0$ between the lines
$nS_1^{(\rm id)}$ and $S_1^{(\rm id)}$.
\label{fig_S0}}
\end{figure}

To describe more precisely the endpoint
$\cE$, we proceed as follows.
We can consider the Wilson line $W_1^{(\mathbf R)}$ and project it parallel onto the
symmetry boundary $\mathbf B$.
The result
is
\be 
n S_1^{(\rm id)} := \overbrace{S_1^{(\rm id)} \oplus \cdots \oplus S_1^{(\rm id)}}^{n \text{ times}} \ ,
\ee 
where $S_1^{(\rm id)}$ is the identity
topological line on $\mathbf B$ and $n$
is the dimension of the representation $\mathbf R$, see Figure \ref{fig_proj}.
(Indeed, due to Dirichlet boundary conditions,
${\rm Tr}_{\mathbf R} \mathrm{Pexp}\int a_1$ collapses to 
${\rm Tr}_{\mathbf R} 
\mathbb I = \dim \mathbf R=n$.)
Then, we can identify the possible endpoints 
$\cE$ as the possible topological junctions
$S_0$ between the lines $n S_1^{(\rm id)}$
and $S_1^{(\rm id)}$ in $\mathbf B$, see Figure \ref{fig_S0}.
The vector space of such topological
junctions is isomorphic to $\mathbb C^n$.
This supports the identification of the possible 
endpoints $\cE$ as the \textit{components}
of an operator transforming
in the representation $\mathbf R$.

Let us now consider an insertion of a
topological operator $S_{d-1}^{(g)}$ inside
$\mathbf B$. 
When we project the bulk
Wilson line $W_1^{(\mathbf R)}$
parallel onto $\mathbf B$,
we obtain a topological point-like
junction $J_0^{(g)}$
between  
$S_{d-1}^{(g)}$ and the line $n S_1^{(\rm id)}$,
see Figure \ref{fig_proj}.
The vector space of such topological
junctions is isomorphic to linear maps
from $\mathbb C^n \rightarrow \mathbb C^n$,
or $n\times n$ complex matrices.
As shown in \cite{Bhardwaj:2023ayw},
these matrices do indeed form a representation
of $G$.
Finally, the action of $S_{d-1}^{(g)}$
on the endpoint $\cE$  corresponds
to fusing $J_0^{(g)}$ and $S_0$,
to obtain a new $S_0'$ junction
between the lines $n S_1^{(\rm id)}$
and $S_1^{(\rm id)}$.
Here $S_0$ and $S_0'$ are identified
with vectors in $\mathbb C^n$,
$J_0^{(g)}$ with an $n\times n$ matrix,
and the fusion implements the matrix
multiplication $S_0' = J_0^{(g)} S_0$.
We have recovered the action of $g\in G$
on the components of the operator transforming in the representation
$\mathbf R$.

\medskip

\noindent{\textbf{A remark on gauging}}. 
In the SymTFT picture for finite symmetries, the gauging operation is obtained via a boundary-changing topological interface, exchanging Dirichlet with Neumann boundary conditions.
Indeed, gauging a finite symmetry is a topological manipulation on the theory. 
In contrast, gauging a 
continuous non-Abelian symmetry
$G^{(0)}$ (including the standard Yang-Mills action and summing/path-integrating over all
bundles and connections) is a non-topological operation.
In the SymTFT 
studied in this work,
one can consider the analogue of a Neumann boundary condition for the global symmetry $G^{(0)}$.
(It might be obstructed if additional bulk terms are present.)
Doing so gives rise to a projector that averages to zero all gauge non-invariant configurations and amounts to summing over flat $G^{(0)}$-connections only. This is distinct from gauging $G^{(0)}$ in a path integral in the standard fashion.
Nonetheless, we expect that
studying obstructions to
topological Neumann boundary conditions for $G^{(0)}$ in the SymTFT can encode information
about obstructions to standard dynamical gauging, i.e.~'t Hooft
anomalies for $G^{(0)}$.

\section{Conclusions and outlook}\label{sec:conclusions}

In this letter we have given a derivation in geometric engineering as well as in holography of the SymTFT for continuous non-Abelian zero-form symmetries. We find the latter is a $d+1$ dimensional Yang-Mills theory at zero coupling, whose flat subsector is dual to the non-Abelian BF theory with the same gauge group. For this reason our results can be seen as a derivation of the proposal in \cite{Antinucci:2024zjp, Brennan:2024fgj}. However, the most convenient duality frame to study this system is provided by the gauge theory side, that can be used to address several key features of the SymTFT. A natural generalization of our findings indicates that for field theories in fewer dimensions, the relevant SymTFT can include topological limits of models that are not just gauge theories. This is the case of 4d SCFTs geometrically engineered in M-theory via conical singularities with $G(2)$-holonomy \cite{Acharya:2023bth}, see also \cite{Braun:2023fqa}. 

Other natural questions that we leave open are in the context of the interplay of our SymTFT with the usual features of ordinary global symmetries, such as anomalies or spontaneous breaking. We plan to return on these topics in future work, thanks to another interesting duality, after 't Hooft 
\cite{tHooft:1981bkw}. Namely one can view the $g_{\rm YM}=0$ limit as a $U(1)^r \rtimes \text{Weyl}(G)$ gauge theory, where $r$ and $\text{Weyl}(G)$ are the rank and Weyl group of $G$. This gives a very explicit way of characterizing the various operators at hand \cite{Cordova:2022rer, Antinucci:2022eat}. In particular, one of the new features of topological defects as symmetries is that they have a higher structure (see e.g.~\cite{Copetti:2023mcq} for a recent discussion). It would be interesting to explore the higher structure of non-Abelian continuous symmetries and its interplay with anomalies from the perspective advocated in this work.

As a final more speculative remark, our findings suggests a pathway to describe topological defects for spacetime symmetries as opposed to internal ones in terms of a gravitational $d+1$ bulk theory at zero coupling.

\phantom{$|$}

\section*{Acknowledgments}

We would like to thank Fabio Apruzzi, Francesco Benini, and Christian Copetti for interesting conversations during the Nordita Program Categorical Aspects of Symmetries,
which inspired this work. We also thank Fabio Apruzzi for interesting discussions at
the 2023 Simons Collaboration on Global Categorical Symmetries Annual Meeting in November related to the projects \cite{Apruzzi:2024htg,DelZotto:2024tae}. We especially thank Kantaro Ohmori for sharing with us his vision and insights. Finally, we are grateful to Greg Moore for taking the time to carefully explain to us   the terminology introduced in \cite{Freed:2022qnc} --- we are honestly relieved that the authors never intended to form sandwiches out of quiches and that was just a misunderstanding. The work of MDZ has received funding from the European Research Council (ERC) under the European Union's Horizon 2020 research and innovation program (grant agreement No. 851931). MDZ also acknowledges support from the Simons Foundation Grant \#888984 (Simons Collaboration on Global Categorical Symmetries), as well as the Swedish Research Council (Grants No. 2022-06593 and 2023-05590). The work of FB is supported by the Simons Collaboration Grant on Global Categorical Symmetries.
The work of RM is partially supported by ERC grants 772408-Stringlandscape and 787320-QBH Structure.

\bibliography{het-PRL}

%merlin.mbs apsrev4-1.bst 2010-07-25 4.21a (PWD, AO, DPC) hacked
%Control: key (0)
%Control: author (72) initials jnrlst
%Control: editor formatted (1) identically to author
%Control: production of article title (-1) disabled
%Control: page (0) single
%Control: year (1) truncated
%Control: production of eprint (0) enabled
\begin{thebibliography}{100}%
\makeatletter
\providecommand \@ifxundefined [1]{%
 \@ifx{#1\undefined}
}%
\providecommand \@ifnum [1]{%
 \ifnum #1\expandafter \@firstoftwo
 \else \expandafter \@secondoftwo
 \fi
}%
\providecommand \@ifx [1]{%
 \ifx #1\expandafter \@firstoftwo
 \else \expandafter \@secondoftwo
 \fi
}%
\providecommand \natexlab [1]{#1}%
\providecommand \enquote  [1]{``#1''}%
\providecommand \bibnamefont  [1]{#1}%
\providecommand \bibfnamefont [1]{#1}%
\providecommand \citenamefont [1]{#1}%
\providecommand \href@noop [0]{\@secondoftwo}%
\providecommand \href [0]{\begingroup \@sanitize@url \@href}%
\providecommand \@href[1]{\@@startlink{#1}\@@href}%
\providecommand \@@href[1]{\endgroup#1\@@endlink}%
\providecommand \@sanitize@url [0]{\catcode `\\12\catcode `\$12\catcode
  `\&12\catcode `\#12\catcode `\^12\catcode `\_12\catcode `\%12\relax}%
\providecommand \@@startlink[1]{}%
\providecommand \@@endlink[0]{}%
\providecommand \url  [0]{\begingroup\@sanitize@url \@url }%
\providecommand \@url [1]{\endgroup\@href {#1}{\urlprefix }}%
\providecommand \urlprefix  [0]{URL }%
\providecommand \Eprint [0]{\href }%
\providecommand \doibase [0]{http://dx.doi.org/}%
\providecommand \selectlanguage [0]{\@gobble}%
\providecommand \bibinfo  [0]{\@secondoftwo}%
\providecommand \bibfield  [0]{\@secondoftwo}%
\providecommand \translation [1]{[#1]}%
\providecommand \BibitemOpen [0]{}%
\providecommand \bibitemStop [0]{}%
\providecommand \bibitemNoStop [0]{.\EOS\space}%
\providecommand \EOS [0]{\spacefactor3000\relax}%
\providecommand \BibitemShut  [1]{\csname bibitem#1\endcsname}%
\let\auto@bib@innerbib\@empty
%</preamble>
\bibitem [{\citenamefont {Gaiotto}\ \emph {et~al.}(2015)\citenamefont
  {Gaiotto}, \citenamefont {Kapustin}, \citenamefont {Seiberg},\ and\
  \citenamefont {Willett}}]{Gaiotto:2014kfa}%
  \BibitemOpen
  \bibfield  {author} {\bibinfo {author} {\bibfnamefont {D.}~\bibnamefont
  {Gaiotto}}, \bibinfo {author} {\bibfnamefont {A.}~\bibnamefont {Kapustin}},
  \bibinfo {author} {\bibfnamefont {N.}~\bibnamefont {Seiberg}}, \ and\
  \bibinfo {author} {\bibfnamefont {B.}~\bibnamefont {Willett}},\ }\href
  {\doibase 10.1007/JHEP02(2015)172} {\bibfield  {journal} {\bibinfo  {journal}
  {JHEP}\ }\textbf {\bibinfo {volume} {02}},\ \bibinfo {pages} {172} (\bibinfo
  {year} {2015})},\ \Eprint {http://arxiv.org/abs/1412.5148} {arXiv:1412.5148
  [hep-th]} \BibitemShut {NoStop}%
\bibitem [{\citenamefont {Cordova}\ \emph
  {et~al.}(2022{\natexlab{a}})\citenamefont {Cordova}, \citenamefont
  {Dumitrescu}, \citenamefont {Intriligator},\ and\ \citenamefont
  {Shao}}]{Cordova:2022ruw}%
  \BibitemOpen
  \bibfield  {author} {\bibinfo {author} {\bibfnamefont {C.}~\bibnamefont
  {Cordova}}, \bibinfo {author} {\bibfnamefont {T.~T.}\ \bibnamefont
  {Dumitrescu}}, \bibinfo {author} {\bibfnamefont {K.}~\bibnamefont
  {Intriligator}}, \ and\ \bibinfo {author} {\bibfnamefont {S.-H.}\
  \bibnamefont {Shao}},\ }in\ \href@noop {} {\emph {\bibinfo {booktitle}
  {{Snowmass 2021}}}}\ (\bibinfo {year} {2022})\ \Eprint
  {http://arxiv.org/abs/2205.09545} {arXiv:2205.09545 [hep-th]} \BibitemShut
  {NoStop}%
\bibitem [{\citenamefont {McGreevy}(2023)}]{McGreevy:2022oyu}%
  \BibitemOpen
  \bibfield  {author} {\bibinfo {author} {\bibfnamefont {J.}~\bibnamefont
  {McGreevy}},\ }\href {\doibase 10.1146/annurev-conmatphys-040721-021029}
  {\bibfield  {journal} {\bibinfo  {journal} {Ann. Rev. Condensed Matter
  Phys.}\ }\textbf {\bibinfo {volume} {14}},\ \bibinfo {pages} {57} (\bibinfo
  {year} {2023})},\ \Eprint {http://arxiv.org/abs/2204.03045} {arXiv:2204.03045
  [cond-mat.str-el]} \BibitemShut {NoStop}%
\bibitem [{\citenamefont {Gomes}(2023)}]{Gomes:2023ahz}%
  \BibitemOpen
  \bibfield  {author} {\bibinfo {author} {\bibfnamefont {P.~R.~S.}\
  \bibnamefont {Gomes}},\ }\href {\doibase 10.21468/SciPostPhysLectNotes.74}
  {\bibfield  {journal} {\bibinfo  {journal} {SciPost Phys. Lect. Notes}\
  }\textbf {\bibinfo {volume} {74}},\ \bibinfo {pages} {1} (\bibinfo {year}
  {2023})},\ \Eprint {http://arxiv.org/abs/2303.01817} {arXiv:2303.01817
  [hep-th]} \BibitemShut {NoStop}%
\bibitem [{\citenamefont {Schafer-Nameki}(2024)}]{Schafer-Nameki:2023jdn}%
  \BibitemOpen
  \bibfield  {author} {\bibinfo {author} {\bibfnamefont {S.}~\bibnamefont
  {Schafer-Nameki}},\ }\href {\doibase 10.1016/j.physrep.2024.01.007}
  {\bibfield  {journal} {\bibinfo  {journal} {Phys. Rept.}\ }\textbf {\bibinfo
  {volume} {1063}},\ \bibinfo {pages} {1} (\bibinfo {year} {2024})},\ \Eprint
  {http://arxiv.org/abs/2305.18296} {arXiv:2305.18296 [hep-th]} \BibitemShut
  {NoStop}%
\bibitem [{\citenamefont {Brennan}\ and\ \citenamefont
  {Hong}(2023)}]{Brennan:2023mmt}%
  \BibitemOpen
  \bibfield  {author} {\bibinfo {author} {\bibfnamefont {T.~D.}\ \bibnamefont
  {Brennan}}\ and\ \bibinfo {author} {\bibfnamefont {S.}~\bibnamefont {Hong}},\
  }\href@noop {} {\  (\bibinfo {year} {2023})},\ \Eprint
  {http://arxiv.org/abs/2306.00912} {arXiv:2306.00912 [hep-ph]} \BibitemShut
  {NoStop}%
\bibitem [{\citenamefont {Bhardwaj}\ \emph
  {et~al.}(2024{\natexlab{a}})\citenamefont {Bhardwaj}, \citenamefont
  {Bottini}, \citenamefont {Fraser-Taliente}, \citenamefont {Gladden},
  \citenamefont {Gould}, \citenamefont {Platschorre},\ and\ \citenamefont
  {Tillim}}]{Bhardwaj:2023kri}%
  \BibitemOpen
  \bibfield  {author} {\bibinfo {author} {\bibfnamefont {L.}~\bibnamefont
  {Bhardwaj}}, \bibinfo {author} {\bibfnamefont {L.~E.}\ \bibnamefont
  {Bottini}}, \bibinfo {author} {\bibfnamefont {L.}~\bibnamefont
  {Fraser-Taliente}}, \bibinfo {author} {\bibfnamefont {L.}~\bibnamefont
  {Gladden}}, \bibinfo {author} {\bibfnamefont {D.~S.~W.}\ \bibnamefont
  {Gould}}, \bibinfo {author} {\bibfnamefont {A.}~\bibnamefont {Platschorre}},
  \ and\ \bibinfo {author} {\bibfnamefont {H.}~\bibnamefont {Tillim}},\ }\href
  {\doibase 10.1016/j.physrep.2023.11.002} {\bibfield  {journal} {\bibinfo
  {journal} {Phys. Rept.}\ }\textbf {\bibinfo {volume} {1051}},\ \bibinfo
  {pages} {1} (\bibinfo {year} {2024}{\natexlab{a}})},\ \Eprint
  {http://arxiv.org/abs/2307.07547} {arXiv:2307.07547 [hep-th]} \BibitemShut
  {NoStop}%
\bibitem [{\citenamefont {Shao}(2023)}]{Shao:2023gho}%
  \BibitemOpen
  \bibfield  {author} {\bibinfo {author} {\bibfnamefont {S.-H.}\ \bibnamefont
  {Shao}},\ }\href@noop {} {\  (\bibinfo {year} {2023})},\ \Eprint
  {http://arxiv.org/abs/2308.00747} {arXiv:2308.00747 [hep-th]} \BibitemShut
  {NoStop}%
\bibitem [{\citenamefont {Carqueville}\ \emph {et~al.}(2023)\citenamefont
  {Carqueville}, \citenamefont {Del~Zotto},\ and\ \citenamefont
  {Runkel}}]{Carqueville:2023jhb}%
  \BibitemOpen
  \bibfield  {author} {\bibinfo {author} {\bibfnamefont {N.}~\bibnamefont
  {Carqueville}}, \bibinfo {author} {\bibfnamefont {M.}~\bibnamefont
  {Del~Zotto}}, \ and\ \bibinfo {author} {\bibfnamefont {I.}~\bibnamefont
  {Runkel}}\ }(\bibinfo {year} {2023})\ \Eprint
  {http://arxiv.org/abs/2311.02449} {arXiv:2311.02449 [math-ph]} \BibitemShut
  {NoStop}%
\bibitem [{\citenamefont {Ji}\ and\ \citenamefont {Wen}(2020)}]{Ji:2019jhk}%
  \BibitemOpen
  \bibfield  {author} {\bibinfo {author} {\bibfnamefont {W.}~\bibnamefont
  {Ji}}\ and\ \bibinfo {author} {\bibfnamefont {X.-G.}\ \bibnamefont {Wen}},\
  }\href {\doibase 10.1103/PhysRevResearch.2.033417} {\bibfield  {journal}
  {\bibinfo  {journal} {Phys. Rev. Res.}\ }\textbf {\bibinfo {volume} {2}},\
  \bibinfo {pages} {033417} (\bibinfo {year} {2020})},\ \Eprint
  {http://arxiv.org/abs/1912.13492} {arXiv:1912.13492 [cond-mat.str-el]}
  \BibitemShut {NoStop}%
\bibitem [{\citenamefont {Gaiotto}\ and\ \citenamefont
  {Kulp}(2021)}]{Gaiotto:2020iye}%
  \BibitemOpen
  \bibfield  {author} {\bibinfo {author} {\bibfnamefont {D.}~\bibnamefont
  {Gaiotto}}\ and\ \bibinfo {author} {\bibfnamefont {J.}~\bibnamefont {Kulp}},\
  }\href {\doibase 10.1007/JHEP02(2021)132} {\bibfield  {journal} {\bibinfo
  {journal} {JHEP}\ }\textbf {\bibinfo {volume} {02}},\ \bibinfo {pages} {132}
  (\bibinfo {year} {2021})},\ \Eprint {http://arxiv.org/abs/2008.05960}
  {arXiv:2008.05960 [hep-th]} \BibitemShut {NoStop}%
\bibitem [{\citenamefont {Apruzzi}\ \emph
  {et~al.}(2023{\natexlab{a}})\citenamefont {Apruzzi}, \citenamefont {Bonetti},
  \citenamefont {Garc{\'\i}a~Etxebarria}, \citenamefont {Hosseini},\ and\
  \citenamefont {Schafer-Nameki}}]{Apruzzi:2021nmk}%
  \BibitemOpen
  \bibfield  {author} {\bibinfo {author} {\bibfnamefont {F.}~\bibnamefont
  {Apruzzi}}, \bibinfo {author} {\bibfnamefont {F.}~\bibnamefont {Bonetti}},
  \bibinfo {author} {\bibfnamefont {I.}~\bibnamefont {Garc{\'\i}a~Etxebarria}},
  \bibinfo {author} {\bibfnamefont {S.~S.}\ \bibnamefont {Hosseini}}, \ and\
  \bibinfo {author} {\bibfnamefont {S.}~\bibnamefont {Schafer-Nameki}},\ }\href
  {\doibase 10.1007/s00220-023-04737-2} {\bibfield  {journal} {\bibinfo
  {journal} {Commun. Math. Phys.}\ }\textbf {\bibinfo {volume} {402}},\
  \bibinfo {pages} {895} (\bibinfo {year} {2023}{\natexlab{a}})},\ \Eprint
  {http://arxiv.org/abs/2112.02092} {arXiv:2112.02092 [hep-th]} \BibitemShut
  {NoStop}%
\bibitem [{\citenamefont {Freed}\ \emph {et~al.}(2022)\citenamefont {Freed},
  \citenamefont {Moore},\ and\ \citenamefont {Teleman}}]{Freed:2022qnc}%
  \BibitemOpen
  \bibfield  {author} {\bibinfo {author} {\bibfnamefont {D.~S.}\ \bibnamefont
  {Freed}}, \bibinfo {author} {\bibfnamefont {G.~W.}\ \bibnamefont {Moore}}, \
  and\ \bibinfo {author} {\bibfnamefont {C.}~\bibnamefont {Teleman}},\
  }\href@noop {} {\  (\bibinfo {year} {2022})},\ \Eprint
  {http://arxiv.org/abs/2209.07471} {arXiv:2209.07471 [hep-th]} \BibitemShut
  {NoStop}%
\bibitem [{\citenamefont {Gukov}\ \emph {et~al.}(2021)\citenamefont {Gukov},
  \citenamefont {Hsin},\ and\ \citenamefont {Pei}}]{Gukov:2020btk}%
  \BibitemOpen
  \bibfield  {author} {\bibinfo {author} {\bibfnamefont {S.}~\bibnamefont
  {Gukov}}, \bibinfo {author} {\bibfnamefont {P.-S.}\ \bibnamefont {Hsin}}, \
  and\ \bibinfo {author} {\bibfnamefont {D.}~\bibnamefont {Pei}},\ }\href
  {\doibase 10.1007/JHEP04(2021)232} {\bibfield  {journal} {\bibinfo  {journal}
  {JHEP}\ }\textbf {\bibinfo {volume} {04}},\ \bibinfo {pages} {232} (\bibinfo
  {year} {2021})},\ \Eprint {http://arxiv.org/abs/2010.15890} {arXiv:2010.15890
  [hep-th]} \BibitemShut {NoStop}%
\bibitem [{\citenamefont {Del~Zotto}\ and\ \citenamefont
  {Garc{\'\i}a~Etxebarria}(2023)}]{DelZotto:2022ras}%
  \BibitemOpen
  \bibfield  {author} {\bibinfo {author} {\bibfnamefont {M.}~\bibnamefont
  {Del~Zotto}}\ and\ \bibinfo {author} {\bibfnamefont {I.}~\bibnamefont
  {Garc{\'\i}a~Etxebarria}},\ }\href {\doibase 10.1007/JHEP11(2023)058}
  {\bibfield  {journal} {\bibinfo  {journal} {JHEP}\ }\textbf {\bibinfo
  {volume} {11}},\ \bibinfo {pages} {058} (\bibinfo {year} {2023})},\ \Eprint
  {http://arxiv.org/abs/2204.06495} {arXiv:2204.06495 [hep-th]} \BibitemShut
  {NoStop}%
\bibitem [{\citenamefont {Bashmakov}\ \emph
  {et~al.}(2023{\natexlab{a}})\citenamefont {Bashmakov}, \citenamefont
  {Del~Zotto},\ and\ \citenamefont {Hasan}}]{Bashmakov:2022jtl}%
  \BibitemOpen
  \bibfield  {author} {\bibinfo {author} {\bibfnamefont {V.}~\bibnamefont
  {Bashmakov}}, \bibinfo {author} {\bibfnamefont {M.}~\bibnamefont
  {Del~Zotto}}, \ and\ \bibinfo {author} {\bibfnamefont {A.}~\bibnamefont
  {Hasan}},\ }\href {\doibase 10.1007/JHEP09(2023)161} {\bibfield  {journal}
  {\bibinfo  {journal} {JHEP}\ }\textbf {\bibinfo {volume} {09}},\ \bibinfo
  {pages} {161} (\bibinfo {year} {2023}{\natexlab{a}})},\ \Eprint
  {http://arxiv.org/abs/2206.07073} {arXiv:2206.07073 [hep-th]} \BibitemShut
  {NoStop}%
\bibitem [{\citenamefont {Kaidi}\ \emph
  {et~al.}(2023{\natexlab{a}})\citenamefont {Kaidi}, \citenamefont {Ohmori},\
  and\ \citenamefont {Zheng}}]{Kaidi:2022cpf}%
  \BibitemOpen
  \bibfield  {author} {\bibinfo {author} {\bibfnamefont {J.}~\bibnamefont
  {Kaidi}}, \bibinfo {author} {\bibfnamefont {K.}~\bibnamefont {Ohmori}}, \
  and\ \bibinfo {author} {\bibfnamefont {Y.}~\bibnamefont {Zheng}},\ }\href
  {\doibase 10.1007/s00220-023-04859-7} {\bibfield  {journal} {\bibinfo
  {journal} {Commun. Math. Phys.}\ }\textbf {\bibinfo {volume} {404}},\
  \bibinfo {pages} {1021} (\bibinfo {year} {2023}{\natexlab{a}})},\ \Eprint
  {http://arxiv.org/abs/2209.11062} {arXiv:2209.11062 [hep-th]} \BibitemShut
  {NoStop}%
\bibitem [{\citenamefont {Bashmakov}\ \emph
  {et~al.}(2023{\natexlab{b}})\citenamefont {Bashmakov}, \citenamefont
  {Del~Zotto}, \citenamefont {Hasan},\ and\ \citenamefont
  {Kaidi}}]{Bashmakov:2022uek}%
  \BibitemOpen
  \bibfield  {author} {\bibinfo {author} {\bibfnamefont {V.}~\bibnamefont
  {Bashmakov}}, \bibinfo {author} {\bibfnamefont {M.}~\bibnamefont
  {Del~Zotto}}, \bibinfo {author} {\bibfnamefont {A.}~\bibnamefont {Hasan}}, \
  and\ \bibinfo {author} {\bibfnamefont {J.}~\bibnamefont {Kaidi}},\ }\href
  {\doibase 10.1007/JHEP05(2023)225} {\bibfield  {journal} {\bibinfo  {journal}
  {JHEP}\ }\textbf {\bibinfo {volume} {05}},\ \bibinfo {pages} {225} (\bibinfo
  {year} {2023}{\natexlab{b}})},\ \Eprint {http://arxiv.org/abs/2211.05138}
  {arXiv:2211.05138 [hep-th]} \BibitemShut {NoStop}%
\bibitem [{\citenamefont {Kaidi}\ \emph
  {et~al.}(2023{\natexlab{b}})\citenamefont {Kaidi}, \citenamefont {Nardoni},
  \citenamefont {Zafrir},\ and\ \citenamefont {Zheng}}]{Kaidi:2023maf}%
  \BibitemOpen
  \bibfield  {author} {\bibinfo {author} {\bibfnamefont {J.}~\bibnamefont
  {Kaidi}}, \bibinfo {author} {\bibfnamefont {E.}~\bibnamefont {Nardoni}},
  \bibinfo {author} {\bibfnamefont {G.}~\bibnamefont {Zafrir}}, \ and\ \bibinfo
  {author} {\bibfnamefont {Y.}~\bibnamefont {Zheng}},\ }\href {\doibase
  10.1007/JHEP10(2023)053} {\bibfield  {journal} {\bibinfo  {journal} {JHEP}\
  }\textbf {\bibinfo {volume} {10}},\ \bibinfo {pages} {053} (\bibinfo {year}
  {2023}{\natexlab{b}})},\ \Eprint {http://arxiv.org/abs/2301.07112}
  {arXiv:2301.07112 [hep-th]} \BibitemShut {NoStop}%
\bibitem [{\citenamefont {Chen}\ \emph {et~al.}(2023)\citenamefont {Chen},
  \citenamefont {Cui}, \citenamefont {Haghighat},\ and\ \citenamefont
  {Wang}}]{Chen:2023qnv}%
  \BibitemOpen
  \bibfield  {author} {\bibinfo {author} {\bibfnamefont {J.}~\bibnamefont
  {Chen}}, \bibinfo {author} {\bibfnamefont {W.}~\bibnamefont {Cui}}, \bibinfo
  {author} {\bibfnamefont {B.}~\bibnamefont {Haghighat}}, \ and\ \bibinfo
  {author} {\bibfnamefont {Y.-N.}\ \bibnamefont {Wang}},\ }\href {\doibase
  10.1007/JHEP11(2023)208} {\bibfield  {journal} {\bibinfo  {journal} {JHEP}\
  }\textbf {\bibinfo {volume} {11}},\ \bibinfo {pages} {208} (\bibinfo {year}
  {2023})},\ \Eprint {http://arxiv.org/abs/2305.09734} {arXiv:2305.09734
  [hep-th]} \BibitemShut {NoStop}%
\bibitem [{\citenamefont {Bashmakov}\ \emph
  {et~al.}(2023{\natexlab{c}})\citenamefont {Bashmakov}, \citenamefont
  {Del~Zotto},\ and\ \citenamefont {Hasan}}]{Bashmakov:2023kwo}%
  \BibitemOpen
  \bibfield  {author} {\bibinfo {author} {\bibfnamefont {V.}~\bibnamefont
  {Bashmakov}}, \bibinfo {author} {\bibfnamefont {M.}~\bibnamefont
  {Del~Zotto}}, \ and\ \bibinfo {author} {\bibfnamefont {A.}~\bibnamefont
  {Hasan}},\ }\href@noop {} {\  (\bibinfo {year} {2023}{\natexlab{c}})},\
  \Eprint {http://arxiv.org/abs/2305.10422} {arXiv:2305.10422 [hep-th]}
  \BibitemShut {NoStop}%
\bibitem [{\citenamefont {Antinucci}\ \emph {et~al.}(2024)\citenamefont
  {Antinucci}, \citenamefont {Copetti}, \citenamefont {Galati},\ and\
  \citenamefont {Rizi}}]{Antinucci:2022cdi}%
  \BibitemOpen
  \bibfield  {author} {\bibinfo {author} {\bibfnamefont {A.}~\bibnamefont
  {Antinucci}}, \bibinfo {author} {\bibfnamefont {C.}~\bibnamefont {Copetti}},
  \bibinfo {author} {\bibfnamefont {G.}~\bibnamefont {Galati}}, \ and\ \bibinfo
  {author} {\bibfnamefont {G.}~\bibnamefont {Rizi}},\ }\href {\doibase
  10.1007/JHEP04(2024)036} {\bibfield  {journal} {\bibinfo  {journal} {JHEP}\
  }\textbf {\bibinfo {volume} {04}},\ \bibinfo {pages} {036} (\bibinfo {year}
  {2024})},\ \Eprint {http://arxiv.org/abs/2212.09549} {arXiv:2212.09549
  [hep-th]} \BibitemShut {NoStop}%
\bibitem [{\citenamefont {Bhardwaj}\ and\ \citenamefont
  {Schafer-Nameki}(2023)}]{Bhardwaj:2023ayw}%
  \BibitemOpen
  \bibfield  {author} {\bibinfo {author} {\bibfnamefont {L.}~\bibnamefont
  {Bhardwaj}}\ and\ \bibinfo {author} {\bibfnamefont {S.}~\bibnamefont
  {Schafer-Nameki}},\ }\href@noop {} {\  (\bibinfo {year} {2023})},\ \Eprint
  {http://arxiv.org/abs/2305.17159} {arXiv:2305.17159 [hep-th]} \BibitemShut
  {NoStop}%
\bibitem [{\citenamefont {Bartsch}\ \emph
  {et~al.}(2023{\natexlab{a}})\citenamefont {Bartsch}, \citenamefont
  {Bullimore},\ and\ \citenamefont {Grigoletto}}]{Bartsch:2023wvv}%
  \BibitemOpen
  \bibfield  {author} {\bibinfo {author} {\bibfnamefont {T.}~\bibnamefont
  {Bartsch}}, \bibinfo {author} {\bibfnamefont {M.}~\bibnamefont {Bullimore}},
  \ and\ \bibinfo {author} {\bibfnamefont {A.}~\bibnamefont {Grigoletto}},\
  }\href@noop {} {\  (\bibinfo {year} {2023}{\natexlab{a}})},\ \Eprint
  {http://arxiv.org/abs/2305.17165} {arXiv:2305.17165 [hep-th]} \BibitemShut
  {NoStop}%
\bibitem [{\citenamefont {Sun}\ and\ \citenamefont
  {Zheng}(2023)}]{Sun:2023xxv}%
  \BibitemOpen
  \bibfield  {author} {\bibinfo {author} {\bibfnamefont {Z.}~\bibnamefont
  {Sun}}\ and\ \bibinfo {author} {\bibfnamefont {Y.}~\bibnamefont {Zheng}},\
  }\href@noop {} {\  (\bibinfo {year} {2023})},\ \Eprint
  {http://arxiv.org/abs/2307.14428} {arXiv:2307.14428 [hep-th]} \BibitemShut
  {NoStop}%
\bibitem [{\citenamefont {Cordova}\ \emph {et~al.}(2024)\citenamefont
  {Cordova}, \citenamefont {Hsin},\ and\ \citenamefont
  {Zhang}}]{Cordova:2023bja}%
  \BibitemOpen
  \bibfield  {author} {\bibinfo {author} {\bibfnamefont {C.}~\bibnamefont
  {Cordova}}, \bibinfo {author} {\bibfnamefont {P.-S.}\ \bibnamefont {Hsin}}, \
  and\ \bibinfo {author} {\bibfnamefont {C.}~\bibnamefont {Zhang}},\ }\href
  {\doibase 10.21468/SciPostPhys.17.5.131} {\bibfield  {journal} {\bibinfo
  {journal} {SciPost Phys.}\ }\textbf {\bibinfo {volume} {17}},\ \bibinfo
  {pages} {131} (\bibinfo {year} {2024})},\ \Eprint
  {http://arxiv.org/abs/2308.11706} {arXiv:2308.11706 [hep-th]} \BibitemShut
  {NoStop}%
\bibitem [{\citenamefont {Antinucci}\ \emph {et~al.}(2023)\citenamefont
  {Antinucci}, \citenamefont {Benini}, \citenamefont {Copetti}, \citenamefont
  {Galati},\ and\ \citenamefont {Rizi}}]{Antinucci:2023ezl}%
  \BibitemOpen
  \bibfield  {author} {\bibinfo {author} {\bibfnamefont {A.}~\bibnamefont
  {Antinucci}}, \bibinfo {author} {\bibfnamefont {F.}~\bibnamefont {Benini}},
  \bibinfo {author} {\bibfnamefont {C.}~\bibnamefont {Copetti}}, \bibinfo
  {author} {\bibfnamefont {G.}~\bibnamefont {Galati}}, \ and\ \bibinfo {author}
  {\bibfnamefont {G.}~\bibnamefont {Rizi}},\ }\href@noop {} {\  (\bibinfo
  {year} {2023})},\ \Eprint {http://arxiv.org/abs/2308.11707} {arXiv:2308.11707
  [hep-th]} \BibitemShut {NoStop}%
\bibitem [{\citenamefont {Bhardwaj}\ \emph
  {et~al.}(2024{\natexlab{b}})\citenamefont {Bhardwaj}, \citenamefont
  {Bottini}, \citenamefont {Pajer},\ and\ \citenamefont
  {Schafer-Nameki}}]{Bhardwaj:2023fca}%
  \BibitemOpen
  \bibfield  {author} {\bibinfo {author} {\bibfnamefont {L.}~\bibnamefont
  {Bhardwaj}}, \bibinfo {author} {\bibfnamefont {L.~E.}\ \bibnamefont
  {Bottini}}, \bibinfo {author} {\bibfnamefont {D.}~\bibnamefont {Pajer}}, \
  and\ \bibinfo {author} {\bibfnamefont {S.}~\bibnamefont {Schafer-Nameki}},\
  }\href {\doibase 10.1103/PhysRevLett.133.161601} {\bibfield  {journal}
  {\bibinfo  {journal} {Phys. Rev. Lett.}\ }\textbf {\bibinfo {volume} {133}},\
  \bibinfo {pages} {161601} (\bibinfo {year} {2024}{\natexlab{b}})},\ \Eprint
  {http://arxiv.org/abs/2310.03786} {arXiv:2310.03786 [cond-mat.str-el]}
  \BibitemShut {NoStop}%
\bibitem [{\citenamefont {Bhardwaj}\ \emph
  {et~al.}(2025{\natexlab{a}})\citenamefont {Bhardwaj}, \citenamefont
  {Bottini}, \citenamefont {Pajer},\ and\ \citenamefont
  {Sch{\"a}fer-Nameki}}]{Bhardwaj:2023idu}%
  \BibitemOpen
  \bibfield  {author} {\bibinfo {author} {\bibfnamefont {L.}~\bibnamefont
  {Bhardwaj}}, \bibinfo {author} {\bibfnamefont {L.~E.}\ \bibnamefont
  {Bottini}}, \bibinfo {author} {\bibfnamefont {D.}~\bibnamefont {Pajer}}, \
  and\ \bibinfo {author} {\bibfnamefont {S.}~\bibnamefont
  {Sch{\"a}fer-Nameki}},\ }\href {\doibase 10.21468/SciPostPhys.18.1.032}
  {\bibfield  {journal} {\bibinfo  {journal} {SciPost Phys.}\ }\textbf
  {\bibinfo {volume} {18}},\ \bibinfo {pages} {032} (\bibinfo {year}
  {2025}{\natexlab{a}})},\ \Eprint {http://arxiv.org/abs/2310.03784}
  {arXiv:2310.03784 [hep-th]} \BibitemShut {NoStop}%
\bibitem [{\citenamefont {Bhardwaj}\ \emph
  {et~al.}(2025{\natexlab{b}})\citenamefont {Bhardwaj}, \citenamefont
  {Bottini}, \citenamefont {Pajer},\ and\ \citenamefont
  {Schafer-Nameki}}]{Bhardwaj:2023bbf}%
  \BibitemOpen
  \bibfield  {author} {\bibinfo {author} {\bibfnamefont {L.}~\bibnamefont
  {Bhardwaj}}, \bibinfo {author} {\bibfnamefont {L.~E.}\ \bibnamefont
  {Bottini}}, \bibinfo {author} {\bibfnamefont {D.}~\bibnamefont {Pajer}}, \
  and\ \bibinfo {author} {\bibfnamefont {S.}~\bibnamefont {Schafer-Nameki}},\
  }\href {\doibase 10.21468/SciPostPhys.18.5.156} {\bibfield  {journal}
  {\bibinfo  {journal} {SciPost Phys.}\ }\textbf {\bibinfo {volume} {18}},\
  \bibinfo {pages} {156} (\bibinfo {year} {2025}{\natexlab{b}})},\ \Eprint
  {http://arxiv.org/abs/2312.17322} {arXiv:2312.17322 [hep-th]} \BibitemShut
  {NoStop}%
\bibitem [{\citenamefont {Brennan}\ and\ \citenamefont
  {Sun}(2024)}]{Brennan:2024fgj}%
  \BibitemOpen
  \bibfield  {author} {\bibinfo {author} {\bibfnamefont {T.~D.}\ \bibnamefont
  {Brennan}}\ and\ \bibinfo {author} {\bibfnamefont {Z.}~\bibnamefont {Sun}},\
  }\href {\doibase 10.1007/JHEP12(2024)100} {\bibfield  {journal} {\bibinfo
  {journal} {JHEP}\ }\textbf {\bibinfo {volume} {12}},\ \bibinfo {pages} {100}
  (\bibinfo {year} {2024})},\ \Eprint {http://arxiv.org/abs/2401.06128}
  {arXiv:2401.06128 [hep-th]} \BibitemShut {NoStop}%
\bibitem [{\citenamefont {Antinucci}\ and\ \citenamefont
  {Benini}(2025)}]{Antinucci:2024zjp}%
  \BibitemOpen
  \bibfield  {author} {\bibinfo {author} {\bibfnamefont {A.}~\bibnamefont
  {Antinucci}}\ and\ \bibinfo {author} {\bibfnamefont {F.}~\bibnamefont
  {Benini}},\ }\href {\doibase 10.1103/PhysRevB.111.024110} {\bibfield
  {journal} {\bibinfo  {journal} {Phys. Rev. B}\ }\textbf {\bibinfo {volume}
  {111}},\ \bibinfo {pages} {024110} (\bibinfo {year} {2025})},\ \Eprint
  {http://arxiv.org/abs/2401.10165} {arXiv:2401.10165 [hep-th]} \BibitemShut
  {NoStop}%
\bibitem [{\citenamefont {Horowitz}(1989)}]{Horowitz:1989ng}%
  \BibitemOpen
  \bibfield  {author} {\bibinfo {author} {\bibfnamefont {G.~T.}\ \bibnamefont
  {Horowitz}},\ }\href {\doibase 10.1007/BF01218410} {\bibfield  {journal}
  {\bibinfo  {journal} {Commun. Math. Phys.}\ }\textbf {\bibinfo {volume}
  {125}},\ \bibinfo {pages} {417} (\bibinfo {year} {1989})}\BibitemShut
  {NoStop}%
\bibitem [{\citenamefont {Apruzzi}\ \emph
  {et~al.}(2024{\natexlab{a}})\citenamefont {Apruzzi}, \citenamefont
  {Bedogna},\ and\ \citenamefont {Dondi}}]{Apruzzi:2024htg}%
  \BibitemOpen
  \bibfield  {author} {\bibinfo {author} {\bibfnamefont {F.}~\bibnamefont
  {Apruzzi}}, \bibinfo {author} {\bibfnamefont {F.}~\bibnamefont {Bedogna}}, \
  and\ \bibinfo {author} {\bibfnamefont {N.}~\bibnamefont {Dondi}},\
  }\href@noop {} {\  (\bibinfo {year} {2024}{\natexlab{a}})},\ \Eprint
  {http://arxiv.org/abs/2402.14813} {arXiv:2402.14813 [hep-th]} \BibitemShut
  {NoStop}%
\bibitem [{\citenamefont {Witten}(1992)}]{Witten:1992xu}%
  \BibitemOpen
  \bibfield  {author} {\bibinfo {author} {\bibfnamefont {E.}~\bibnamefont
  {Witten}},\ }\href {\doibase 10.1016/0393-0440(92)90034-X} {\bibfield
  {journal} {\bibinfo  {journal} {J. Geom. Phys.}\ }\textbf {\bibinfo {volume}
  {9}},\ \bibinfo {pages} {303} (\bibinfo {year} {1992})},\ \Eprint
  {http://arxiv.org/abs/hep-th/9204083} {arXiv:hep-th/9204083} \BibitemShut
  {NoStop}%
\bibitem [{\citenamefont {Cordova}\ \emph
  {et~al.}(2022{\natexlab{b}})\citenamefont {Cordova}, \citenamefont {Ohmori},\
  and\ \citenamefont {Rudelius}}]{Cordova:2022rer}%
  \BibitemOpen
  \bibfield  {author} {\bibinfo {author} {\bibfnamefont {C.}~\bibnamefont
  {Cordova}}, \bibinfo {author} {\bibfnamefont {K.}~\bibnamefont {Ohmori}}, \
  and\ \bibinfo {author} {\bibfnamefont {T.}~\bibnamefont {Rudelius}},\ }\href
  {\doibase 10.1007/JHEP11(2022)154} {\bibfield  {journal} {\bibinfo  {journal}
  {JHEP}\ }\textbf {\bibinfo {volume} {11}},\ \bibinfo {pages} {154} (\bibinfo
  {year} {2022}{\natexlab{b}})},\ \Eprint {http://arxiv.org/abs/2202.05866}
  {arXiv:2202.05866 [hep-th]} \BibitemShut {NoStop}%
\bibitem [{\citenamefont {Antinucci}\ \emph {et~al.}(2022)\citenamefont
  {Antinucci}, \citenamefont {Galati},\ and\ \citenamefont
  {Rizi}}]{Antinucci:2022eat}%
  \BibitemOpen
  \bibfield  {author} {\bibinfo {author} {\bibfnamefont {A.}~\bibnamefont
  {Antinucci}}, \bibinfo {author} {\bibfnamefont {G.}~\bibnamefont {Galati}}, \
  and\ \bibinfo {author} {\bibfnamefont {G.}~\bibnamefont {Rizi}},\ }\href
  {\doibase 10.1007/JHEP12(2022)061} {\bibfield  {journal} {\bibinfo  {journal}
  {JHEP}\ }\textbf {\bibinfo {volume} {12}},\ \bibinfo {pages} {061} (\bibinfo
  {year} {2022})},\ \Eprint {http://arxiv.org/abs/2206.05646} {arXiv:2206.05646
  [hep-th]} \BibitemShut {NoStop}%
\bibitem [{\citenamefont {Xie}\ and\ \citenamefont {Yau}(2017)}]{Xie:2017pfl}%
  \BibitemOpen
  \bibfield  {author} {\bibinfo {author} {\bibfnamefont {D.}~\bibnamefont
  {Xie}}\ and\ \bibinfo {author} {\bibfnamefont {S.-T.}\ \bibnamefont {Yau}},\
  }\href {\doibase 10.1007/JHEP06(2017)134} {\bibfield  {journal} {\bibinfo
  {journal} {JHEP}\ }\textbf {\bibinfo {volume} {06}},\ \bibinfo {pages} {134}
  (\bibinfo {year} {2017})},\ \Eprint {http://arxiv.org/abs/1704.00799}
  {arXiv:1704.00799 [hep-th]} \BibitemShut {NoStop}%
\bibitem [{\citenamefont {Acharya}\ \emph {et~al.}(2022)\citenamefont
  {Acharya}, \citenamefont {Lambert}, \citenamefont {Najjar}, \citenamefont
  {Svanes},\ and\ \citenamefont {Tian}}]{Acharya:2021jsp}%
  \BibitemOpen
  \bibfield  {author} {\bibinfo {author} {\bibfnamefont {B.}~\bibnamefont
  {Acharya}}, \bibinfo {author} {\bibfnamefont {N.}~\bibnamefont {Lambert}},
  \bibinfo {author} {\bibfnamefont {M.}~\bibnamefont {Najjar}}, \bibinfo
  {author} {\bibfnamefont {E.~E.}\ \bibnamefont {Svanes}}, \ and\ \bibinfo
  {author} {\bibfnamefont {J.}~\bibnamefont {Tian}},\ }\href {\doibase
  10.1007/JHEP04(2022)114} {\bibfield  {journal} {\bibinfo  {journal} {JHEP}\
  }\textbf {\bibinfo {volume} {04}},\ \bibinfo {pages} {114} (\bibinfo {year}
  {2022})},\ \Eprint {http://arxiv.org/abs/2110.14441} {arXiv:2110.14441
  [hep-th]} \BibitemShut {NoStop}%
\bibitem [{\citenamefont {Tian}\ and\ \citenamefont
  {Wang}(2022)}]{Tian:2021cif}%
  \BibitemOpen
  \bibfield  {author} {\bibinfo {author} {\bibfnamefont {J.}~\bibnamefont
  {Tian}}\ and\ \bibinfo {author} {\bibfnamefont {Y.-N.}\ \bibnamefont
  {Wang}},\ }\href {\doibase 10.21468/SciPostPhys.12.4.127} {\bibfield
  {journal} {\bibinfo  {journal} {SciPost Phys.}\ }\textbf {\bibinfo {volume}
  {12}},\ \bibinfo {pages} {127} (\bibinfo {year} {2022})},\ \Eprint
  {http://arxiv.org/abs/2110.15129} {arXiv:2110.15129 [hep-th]} \BibitemShut
  {NoStop}%
\bibitem [{\citenamefont {Del~Zotto}\ \emph
  {et~al.}(2022{\natexlab{a}})\citenamefont {Del~Zotto}, \citenamefont
  {Heckman}, \citenamefont {Meynet}, \citenamefont {Moscrop},\ and\
  \citenamefont {Zhang}}]{DelZotto:2022fnw}%
  \BibitemOpen
  \bibfield  {author} {\bibinfo {author} {\bibfnamefont {M.}~\bibnamefont
  {Del~Zotto}}, \bibinfo {author} {\bibfnamefont {J.~J.}\ \bibnamefont
  {Heckman}}, \bibinfo {author} {\bibfnamefont {S.~N.}\ \bibnamefont {Meynet}},
  \bibinfo {author} {\bibfnamefont {R.}~\bibnamefont {Moscrop}}, \ and\
  \bibinfo {author} {\bibfnamefont {H.~Y.}\ \bibnamefont {Zhang}},\ }\href
  {\doibase 10.1103/PhysRevD.106.046010} {\bibfield  {journal} {\bibinfo
  {journal} {Phys. Rev. D}\ }\textbf {\bibinfo {volume} {106}},\ \bibinfo
  {pages} {046010} (\bibinfo {year} {2022}{\natexlab{a}})},\ \Eprint
  {http://arxiv.org/abs/2201.08372} {arXiv:2201.08372 [hep-th]} \BibitemShut
  {NoStop}%
\bibitem [{\citenamefont {van Beest}\ \emph {et~al.}(2023)\citenamefont {van
  Beest}, \citenamefont {Gould}, \citenamefont {Schafer-Nameki},\ and\
  \citenamefont {Wang}}]{vanBeest:2022fss}%
  \BibitemOpen
  \bibfield  {author} {\bibinfo {author} {\bibfnamefont {M.}~\bibnamefont {van
  Beest}}, \bibinfo {author} {\bibfnamefont {D.~S.~W.}\ \bibnamefont {Gould}},
  \bibinfo {author} {\bibfnamefont {S.}~\bibnamefont {Schafer-Nameki}}, \ and\
  \bibinfo {author} {\bibfnamefont {Y.-N.}\ \bibnamefont {Wang}},\ }\href
  {\doibase 10.1007/JHEP02(2023)226} {\bibfield  {journal} {\bibinfo  {journal}
  {JHEP}\ }\textbf {\bibinfo {volume} {02}},\ \bibinfo {pages} {226} (\bibinfo
  {year} {2023})},\ \Eprint {http://arxiv.org/abs/2210.03703} {arXiv:2210.03703
  [hep-th]} \BibitemShut {NoStop}%
\bibitem [{\citenamefont {Apruzzi}\ \emph
  {et~al.}(2023{\natexlab{b}})\citenamefont {Apruzzi}, \citenamefont {Bah},
  \citenamefont {Bonetti},\ and\ \citenamefont
  {Schafer-Nameki}}]{Apruzzi:2022rei}%
  \BibitemOpen
  \bibfield  {author} {\bibinfo {author} {\bibfnamefont {F.}~\bibnamefont
  {Apruzzi}}, \bibinfo {author} {\bibfnamefont {I.}~\bibnamefont {Bah}},
  \bibinfo {author} {\bibfnamefont {F.}~\bibnamefont {Bonetti}}, \ and\
  \bibinfo {author} {\bibfnamefont {S.}~\bibnamefont {Schafer-Nameki}},\ }\href
  {\doibase 10.1103/PhysRevLett.130.121601} {\bibfield  {journal} {\bibinfo
  {journal} {Phys. Rev. Lett.}\ }\textbf {\bibinfo {volume} {130}},\ \bibinfo
  {pages} {121601} (\bibinfo {year} {2023}{\natexlab{b}})},\ \Eprint
  {http://arxiv.org/abs/2208.07373} {arXiv:2208.07373 [hep-th]} \BibitemShut
  {NoStop}%
\bibitem [{\citenamefont
  {Garc{\'\i}a~Etxebarria}(2022)}]{GarciaEtxebarria:2022vzq}%
  \BibitemOpen
  \bibfield  {author} {\bibinfo {author} {\bibfnamefont {I.}~\bibnamefont
  {Garc{\'\i}a~Etxebarria}},\ }\href {\doibase 10.1002/prop.202200154}
  {\bibfield  {journal} {\bibinfo  {journal} {Fortsch. Phys.}\ }\textbf
  {\bibinfo {volume} {70}},\ \bibinfo {pages} {2200154} (\bibinfo {year}
  {2022})},\ \Eprint {http://arxiv.org/abs/2208.07508} {arXiv:2208.07508
  [hep-th]} \BibitemShut {NoStop}%
\bibitem [{\citenamefont {Heckman}\ \emph
  {et~al.}(2023{\natexlab{a}})\citenamefont {Heckman}, \citenamefont
  {H{\"u}bner}, \citenamefont {Torres},\ and\ \citenamefont
  {Zhang}}]{Heckman:2022muc}%
  \BibitemOpen
  \bibfield  {author} {\bibinfo {author} {\bibfnamefont {J.~J.}\ \bibnamefont
  {Heckman}}, \bibinfo {author} {\bibfnamefont {M.}~\bibnamefont {H{\"u}bner}},
  \bibinfo {author} {\bibfnamefont {E.}~\bibnamefont {Torres}}, \ and\ \bibinfo
  {author} {\bibfnamefont {H.~Y.}\ \bibnamefont {Zhang}},\ }\href {\doibase
  10.1002/prop.202200180} {\bibfield  {journal} {\bibinfo  {journal} {Fortsch.
  Phys.}\ }\textbf {\bibinfo {volume} {71}},\ \bibinfo {pages} {2200180}
  (\bibinfo {year} {2023}{\natexlab{a}})},\ \Eprint
  {http://arxiv.org/abs/2209.03343} {arXiv:2209.03343 [hep-th]} \BibitemShut
  {NoStop}%
\bibitem [{\citenamefont {Heckman}\ \emph
  {et~al.}(2023{\natexlab{b}})\citenamefont {Heckman}, \citenamefont {Hubner},
  \citenamefont {Torres}, \citenamefont {Yu},\ and\ \citenamefont
  {Zhang}}]{Heckman:2022xgu}%
  \BibitemOpen
  \bibfield  {author} {\bibinfo {author} {\bibfnamefont {J.~J.}\ \bibnamefont
  {Heckman}}, \bibinfo {author} {\bibfnamefont {M.}~\bibnamefont {Hubner}},
  \bibinfo {author} {\bibfnamefont {E.}~\bibnamefont {Torres}}, \bibinfo
  {author} {\bibfnamefont {X.}~\bibnamefont {Yu}}, \ and\ \bibinfo {author}
  {\bibfnamefont {H.~Y.}\ \bibnamefont {Zhang}},\ }\href {\doibase
  10.1103/PhysRevD.108.046015} {\bibfield  {journal} {\bibinfo  {journal}
  {Phys. Rev. D}\ }\textbf {\bibinfo {volume} {108}},\ \bibinfo {pages}
  {046015} (\bibinfo {year} {2023}{\natexlab{b}})},\ \Eprint
  {http://arxiv.org/abs/2212.09743} {arXiv:2212.09743 [hep-th]} \BibitemShut
  {NoStop}%
\bibitem [{\citenamefont {Etheredge}\ \emph {et~al.}(2023)\citenamefont
  {Etheredge}, \citenamefont {Garcia~Etxebarria}, \citenamefont {Heidenreich},\
  and\ \citenamefont {Rauch}}]{Etheredge:2023ler}%
  \BibitemOpen
  \bibfield  {author} {\bibinfo {author} {\bibfnamefont {M.}~\bibnamefont
  {Etheredge}}, \bibinfo {author} {\bibfnamefont {I.}~\bibnamefont
  {Garcia~Etxebarria}}, \bibinfo {author} {\bibfnamefont {B.}~\bibnamefont
  {Heidenreich}}, \ and\ \bibinfo {author} {\bibfnamefont {S.}~\bibnamefont
  {Rauch}},\ }\href {\doibase 10.1007/JHEP09(2023)005} {\bibfield  {journal}
  {\bibinfo  {journal} {JHEP}\ }\textbf {\bibinfo {volume} {09}},\ \bibinfo
  {pages} {005} (\bibinfo {year} {2023})},\ \Eprint
  {http://arxiv.org/abs/2302.14068} {arXiv:2302.14068 [hep-th]} \BibitemShut
  {NoStop}%
\bibitem [{\citenamefont {Dierigl}\ \emph {et~al.}(2024)\citenamefont
  {Dierigl}, \citenamefont {Heckman}, \citenamefont {Montero},\ and\
  \citenamefont {Torres}}]{Dierigl:2023jdp}%
  \BibitemOpen
  \bibfield  {author} {\bibinfo {author} {\bibfnamefont {M.}~\bibnamefont
  {Dierigl}}, \bibinfo {author} {\bibfnamefont {J.~J.}\ \bibnamefont
  {Heckman}}, \bibinfo {author} {\bibfnamefont {M.}~\bibnamefont {Montero}}, \
  and\ \bibinfo {author} {\bibfnamefont {E.}~\bibnamefont {Torres}},\ }\href
  {\doibase 10.1103/PhysRevD.109.046004} {\bibfield  {journal} {\bibinfo
  {journal} {Phys. Rev. D}\ }\textbf {\bibinfo {volume} {109}},\ \bibinfo
  {pages} {046004} (\bibinfo {year} {2024})},\ \Eprint
  {http://arxiv.org/abs/2305.05689} {arXiv:2305.05689 [hep-th]} \BibitemShut
  {NoStop}%
\bibitem [{\citenamefont {Cveti{\v{c}}}\ \emph {et~al.}(2024)\citenamefont
  {Cveti{\v{c}}}, \citenamefont {Heckman}, \citenamefont {H{\"u}bner},\ and\
  \citenamefont {Torres}}]{Cvetic:2023plv}%
  \BibitemOpen
  \bibfield  {author} {\bibinfo {author} {\bibfnamefont {M.}~\bibnamefont
  {Cveti{\v{c}}}}, \bibinfo {author} {\bibfnamefont {J.~J.}\ \bibnamefont
  {Heckman}}, \bibinfo {author} {\bibfnamefont {M.}~\bibnamefont {H{\"u}bner}},
  \ and\ \bibinfo {author} {\bibfnamefont {E.}~\bibnamefont {Torres}},\ }\href
  {\doibase 10.1103/PhysRevD.109.046007} {\bibfield  {journal} {\bibinfo
  {journal} {Phys. Rev. D}\ }\textbf {\bibinfo {volume} {109}},\ \bibinfo
  {pages} {046007} (\bibinfo {year} {2024})},\ \Eprint
  {http://arxiv.org/abs/2305.09665} {arXiv:2305.09665 [hep-th]} \BibitemShut
  {NoStop}%
\bibitem [{\citenamefont {Lawrie}\ \emph {et~al.}(2024)\citenamefont {Lawrie},
  \citenamefont {Yu},\ and\ \citenamefont {Zhang}}]{Lawrie:2023tdz}%
  \BibitemOpen
  \bibfield  {author} {\bibinfo {author} {\bibfnamefont {C.}~\bibnamefont
  {Lawrie}}, \bibinfo {author} {\bibfnamefont {X.}~\bibnamefont {Yu}}, \ and\
  \bibinfo {author} {\bibfnamefont {H.~Y.}\ \bibnamefont {Zhang}},\ }\href
  {\doibase 10.1103/PhysRevD.109.026005} {\bibfield  {journal} {\bibinfo
  {journal} {Phys. Rev. D}\ }\textbf {\bibinfo {volume} {109}},\ \bibinfo
  {pages} {026005} (\bibinfo {year} {2024})},\ \Eprint
  {http://arxiv.org/abs/2306.11783} {arXiv:2306.11783 [hep-th]} \BibitemShut
  {NoStop}%
\bibitem [{\citenamefont {Bah}\ \emph {et~al.}(2024)\citenamefont {Bah},
  \citenamefont {Leung},\ and\ \citenamefont {Waddleton}}]{Bah:2023ymy}%
  \BibitemOpen
  \bibfield  {author} {\bibinfo {author} {\bibfnamefont {I.}~\bibnamefont
  {Bah}}, \bibinfo {author} {\bibfnamefont {E.}~\bibnamefont {Leung}}, \ and\
  \bibinfo {author} {\bibfnamefont {T.}~\bibnamefont {Waddleton}},\ }\href
  {\doibase 10.1007/JHEP01(2024)117} {\bibfield  {journal} {\bibinfo  {journal}
  {JHEP}\ }\textbf {\bibinfo {volume} {01}},\ \bibinfo {pages} {117} (\bibinfo
  {year} {2024})},\ \Eprint {http://arxiv.org/abs/2306.15783} {arXiv:2306.15783
  [hep-th]} \BibitemShut {NoStop}%
\bibitem [{\citenamefont {Apruzzi}\ \emph
  {et~al.}(2024{\natexlab{b}})\citenamefont {Apruzzi}, \citenamefont {Bonetti},
  \citenamefont {Gould},\ and\ \citenamefont
  {Schafer-Nameki}}]{Apruzzi:2023uma}%
  \BibitemOpen
  \bibfield  {author} {\bibinfo {author} {\bibfnamefont {F.}~\bibnamefont
  {Apruzzi}}, \bibinfo {author} {\bibfnamefont {F.}~\bibnamefont {Bonetti}},
  \bibinfo {author} {\bibfnamefont {D.~S.~W.}\ \bibnamefont {Gould}}, \ and\
  \bibinfo {author} {\bibfnamefont {S.}~\bibnamefont {Schafer-Nameki}},\ }\href
  {\doibase 10.21468/SciPostPhys.17.1.025} {\bibfield  {journal} {\bibinfo
  {journal} {SciPost Phys.}\ }\textbf {\bibinfo {volume} {17}},\ \bibinfo
  {pages} {025} (\bibinfo {year} {2024}{\natexlab{b}})},\ \Eprint
  {http://arxiv.org/abs/2306.16405} {arXiv:2306.16405 [hep-th]} \BibitemShut
  {NoStop}%
\bibitem [{\citenamefont {Baume}\ \emph {et~al.}(2024)\citenamefont {Baume},
  \citenamefont {Heckman}, \citenamefont {H{\"u}bner}, \citenamefont {Torres},
  \citenamefont {Turner},\ and\ \citenamefont {Yu}}]{Baume:2023kkf}%
  \BibitemOpen
  \bibfield  {author} {\bibinfo {author} {\bibfnamefont {F.}~\bibnamefont
  {Baume}}, \bibinfo {author} {\bibfnamefont {J.~J.}\ \bibnamefont {Heckman}},
  \bibinfo {author} {\bibfnamefont {M.}~\bibnamefont {H{\"u}bner}}, \bibinfo
  {author} {\bibfnamefont {E.}~\bibnamefont {Torres}}, \bibinfo {author}
  {\bibfnamefont {A.~P.}\ \bibnamefont {Turner}}, \ and\ \bibinfo {author}
  {\bibfnamefont {X.}~\bibnamefont {Yu}},\ }\href {\doibase
  10.1103/PhysRevD.109.106013} {\bibfield  {journal} {\bibinfo  {journal}
  {Phys. Rev. D}\ }\textbf {\bibinfo {volume} {109}},\ \bibinfo {pages}
  {106013} (\bibinfo {year} {2024})},\ \Eprint
  {http://arxiv.org/abs/2310.12980} {arXiv:2310.12980 [hep-th]} \BibitemShut
  {NoStop}%
\bibitem [{\citenamefont {Yu}(2024)}]{Yu:2023nyn}%
  \BibitemOpen
  \bibfield  {author} {\bibinfo {author} {\bibfnamefont {X.}~\bibnamefont
  {Yu}},\ }\href {\doibase 10.1103/PhysRevD.110.065008} {\bibfield  {journal}
  {\bibinfo  {journal} {Phys. Rev. D}\ }\textbf {\bibinfo {volume} {110}},\
  \bibinfo {pages} {065008} (\bibinfo {year} {2024})},\ \Eprint
  {http://arxiv.org/abs/2310.15339} {arXiv:2310.15339 [hep-th]} \BibitemShut
  {NoStop}%
\bibitem [{\citenamefont {Heckman}\ \emph
  {et~al.}(2024{\natexlab{a}})\citenamefont {Heckman}, \citenamefont
  {H{\"u}bner},\ and\ \citenamefont {Murdia}}]{Heckman:2024oot}%
  \BibitemOpen
  \bibfield  {author} {\bibinfo {author} {\bibfnamefont {J.~J.}\ \bibnamefont
  {Heckman}}, \bibinfo {author} {\bibfnamefont {M.}~\bibnamefont {H{\"u}bner}},
  \ and\ \bibinfo {author} {\bibfnamefont {C.}~\bibnamefont {Murdia}},\ }\href
  {\doibase 10.1103/PhysRevD.110.046007} {\bibfield  {journal} {\bibinfo
  {journal} {Phys. Rev. D}\ }\textbf {\bibinfo {volume} {110}},\ \bibinfo
  {pages} {046007} (\bibinfo {year} {2024}{\natexlab{a}})},\ \Eprint
  {http://arxiv.org/abs/2401.09538} {arXiv:2401.09538 [hep-th]} \BibitemShut
  {NoStop}%
\bibitem [{\citenamefont {Heckman}\ \emph
  {et~al.}(2024{\natexlab{b}})\citenamefont {Heckman}, \citenamefont
  {McNamara}, \citenamefont {Montero}, \citenamefont {Sharon}, \citenamefont
  {Vafa},\ and\ \citenamefont {Valenzuela}}]{Heckman:2024obe}%
  \BibitemOpen
  \bibfield  {author} {\bibinfo {author} {\bibfnamefont {J.~J.}\ \bibnamefont
  {Heckman}}, \bibinfo {author} {\bibfnamefont {J.}~\bibnamefont {McNamara}},
  \bibinfo {author} {\bibfnamefont {M.}~\bibnamefont {Montero}}, \bibinfo
  {author} {\bibfnamefont {A.}~\bibnamefont {Sharon}}, \bibinfo {author}
  {\bibfnamefont {C.}~\bibnamefont {Vafa}}, \ and\ \bibinfo {author}
  {\bibfnamefont {I.}~\bibnamefont {Valenzuela}},\ }\href {\doibase
  10.1103/PhysRevD.110.106001} {\bibfield  {journal} {\bibinfo  {journal}
  {Phys. Rev. D}\ }\textbf {\bibinfo {volume} {110}},\ \bibinfo {pages}
  {106001} (\bibinfo {year} {2024}{\natexlab{b}})},\ \Eprint
  {http://arxiv.org/abs/2402.00118} {arXiv:2402.00118 [hep-th]} \BibitemShut
  {NoStop}%
\bibitem [{\citenamefont {Del~Zotto}\ \emph {et~al.}(2024)\citenamefont
  {Del~Zotto}, \citenamefont {Meynet},\ and\ \citenamefont
  {Moscrop}}]{DelZotto:2024tae}%
  \BibitemOpen
  \bibfield  {author} {\bibinfo {author} {\bibfnamefont {M.}~\bibnamefont
  {Del~Zotto}}, \bibinfo {author} {\bibfnamefont {S.~N.}\ \bibnamefont
  {Meynet}}, \ and\ \bibinfo {author} {\bibfnamefont {R.}~\bibnamefont
  {Moscrop}},\ }\href {\doibase 10.1007/JHEP07(2024)220} {\bibfield  {journal}
  {\bibinfo  {journal} {JHEP}\ }\textbf {\bibinfo {volume} {07}},\ \bibinfo
  {pages} {220} (\bibinfo {year} {2024})},\ \Eprint
  {http://arxiv.org/abs/2402.18646} {arXiv:2402.18646 [hep-th]} \BibitemShut
  {NoStop}%
\bibitem [{\citenamefont {Del~Zotto}\ \emph {et~al.}(2016)\citenamefont
  {Del~Zotto}, \citenamefont {Heckman}, \citenamefont {Park},\ and\
  \citenamefont {Rudelius}}]{DelZotto:2015isa}%
  \BibitemOpen
  \bibfield  {author} {\bibinfo {author} {\bibfnamefont {M.}~\bibnamefont
  {Del~Zotto}}, \bibinfo {author} {\bibfnamefont {J.~J.}\ \bibnamefont
  {Heckman}}, \bibinfo {author} {\bibfnamefont {D.~S.}\ \bibnamefont {Park}}, \
  and\ \bibinfo {author} {\bibfnamefont {T.}~\bibnamefont {Rudelius}},\ }\href
  {\doibase 10.1007/s11005-016-0839-5} {\bibfield  {journal} {\bibinfo
  {journal} {Lett. Math. Phys.}\ }\textbf {\bibinfo {volume} {106}},\ \bibinfo
  {pages} {765} (\bibinfo {year} {2016})},\ \Eprint
  {http://arxiv.org/abs/1503.04806} {arXiv:1503.04806 [hep-th]} \BibitemShut
  {NoStop}%
\bibitem [{\citenamefont {Garc{\'\i}a~Etxebarria}\ \emph
  {et~al.}(2019)\citenamefont {Garc{\'\i}a~Etxebarria}, \citenamefont
  {Heidenreich},\ and\ \citenamefont {Regalado}}]{GarciaEtxebarria:2019caf}%
  \BibitemOpen
  \bibfield  {author} {\bibinfo {author} {\bibfnamefont {I.}~\bibnamefont
  {Garc{\'\i}a~Etxebarria}}, \bibinfo {author} {\bibfnamefont {B.}~\bibnamefont
  {Heidenreich}}, \ and\ \bibinfo {author} {\bibfnamefont {D.}~\bibnamefont
  {Regalado}},\ }\href {\doibase 10.1007/JHEP10(2019)169} {\bibfield  {journal}
  {\bibinfo  {journal} {JHEP}\ }\textbf {\bibinfo {volume} {10}},\ \bibinfo
  {pages} {169} (\bibinfo {year} {2019})},\ \Eprint
  {http://arxiv.org/abs/1908.08027} {arXiv:1908.08027 [hep-th]} \BibitemShut
  {NoStop}%
\bibitem [{\citenamefont {Morrison}\ \emph {et~al.}(2020)\citenamefont
  {Morrison}, \citenamefont {Schafer-Nameki},\ and\ \citenamefont
  {Willett}}]{Morrison:2020ool}%
  \BibitemOpen
  \bibfield  {author} {\bibinfo {author} {\bibfnamefont {D.~R.}\ \bibnamefont
  {Morrison}}, \bibinfo {author} {\bibfnamefont {S.}~\bibnamefont
  {Schafer-Nameki}}, \ and\ \bibinfo {author} {\bibfnamefont {B.}~\bibnamefont
  {Willett}},\ }\href {\doibase 10.1007/JHEP09(2020)024} {\bibfield  {journal}
  {\bibinfo  {journal} {JHEP}\ }\textbf {\bibinfo {volume} {09}},\ \bibinfo
  {pages} {024} (\bibinfo {year} {2020})},\ \Eprint
  {http://arxiv.org/abs/2005.12296} {arXiv:2005.12296 [hep-th]} \BibitemShut
  {NoStop}%
\bibitem [{\citenamefont {Albertini}\ \emph {et~al.}(2020)\citenamefont
  {Albertini}, \citenamefont {Del~Zotto}, \citenamefont
  {Garc{\'\i}a~Etxebarria},\ and\ \citenamefont
  {Hosseini}}]{Albertini:2020mdx}%
  \BibitemOpen
  \bibfield  {author} {\bibinfo {author} {\bibfnamefont {F.}~\bibnamefont
  {Albertini}}, \bibinfo {author} {\bibfnamefont {M.}~\bibnamefont
  {Del~Zotto}}, \bibinfo {author} {\bibfnamefont {I.}~\bibnamefont
  {Garc{\'\i}a~Etxebarria}}, \ and\ \bibinfo {author} {\bibfnamefont {S.~S.}\
  \bibnamefont {Hosseini}},\ }\href {\doibase 10.1007/JHEP12(2020)203}
  {\bibfield  {journal} {\bibinfo  {journal} {JHEP}\ }\textbf {\bibinfo
  {volume} {12}},\ \bibinfo {pages} {203} (\bibinfo {year} {2020})},\ \Eprint
  {http://arxiv.org/abs/2005.12831} {arXiv:2005.12831 [hep-th]} \BibitemShut
  {NoStop}%
\bibitem [{\citenamefont {Hubner}\ \emph {et~al.}(2022)\citenamefont {Hubner},
  \citenamefont {Morrison}, \citenamefont {Schafer-Nameki},\ and\ \citenamefont
  {Wang}}]{Hubner:2022kxr}%
  \BibitemOpen
  \bibfield  {author} {\bibinfo {author} {\bibfnamefont {M.}~\bibnamefont
  {Hubner}}, \bibinfo {author} {\bibfnamefont {D.~R.}\ \bibnamefont
  {Morrison}}, \bibinfo {author} {\bibfnamefont {S.}~\bibnamefont
  {Schafer-Nameki}}, \ and\ \bibinfo {author} {\bibfnamefont {Y.-N.}\
  \bibnamefont {Wang}},\ }\href {\doibase 10.21468/SciPostPhys.13.2.030}
  {\bibfield  {journal} {\bibinfo  {journal} {SciPost Phys.}\ }\textbf
  {\bibinfo {volume} {13}},\ \bibinfo {pages} {030} (\bibinfo {year} {2022})},\
  \Eprint {http://arxiv.org/abs/2203.10022} {arXiv:2203.10022 [hep-th]}
  \BibitemShut {NoStop}%
\bibitem [{\citenamefont {Benini}\ \emph {et~al.}(2009)\citenamefont {Benini},
  \citenamefont {Benvenuti},\ and\ \citenamefont {Tachikawa}}]{Benini:2009gi}%
  \BibitemOpen
  \bibfield  {author} {\bibinfo {author} {\bibfnamefont {F.}~\bibnamefont
  {Benini}}, \bibinfo {author} {\bibfnamefont {S.}~\bibnamefont {Benvenuti}}, \
  and\ \bibinfo {author} {\bibfnamefont {Y.}~\bibnamefont {Tachikawa}},\ }\href
  {\doibase 10.1088/1126-6708/2009/09/052} {\bibfield  {journal} {\bibinfo
  {journal} {JHEP}\ }\textbf {\bibinfo {volume} {09}},\ \bibinfo {pages} {052}
  (\bibinfo {year} {2009})},\ \Eprint {http://arxiv.org/abs/0906.0359}
  {arXiv:0906.0359 [hep-th]} \BibitemShut {NoStop}%
\bibitem [{\citenamefont {Del~Zotto}\ \emph {et~al.}(2015)\citenamefont
  {Del~Zotto}, \citenamefont {Heckman}, \citenamefont {Tomasiello},\ and\
  \citenamefont {Vafa}}]{DelZotto:2014hpa}%
  \BibitemOpen
  \bibfield  {author} {\bibinfo {author} {\bibfnamefont {M.}~\bibnamefont
  {Del~Zotto}}, \bibinfo {author} {\bibfnamefont {J.~J.}\ \bibnamefont
  {Heckman}}, \bibinfo {author} {\bibfnamefont {A.}~\bibnamefont {Tomasiello}},
  \ and\ \bibinfo {author} {\bibfnamefont {C.}~\bibnamefont {Vafa}},\ }\href
  {\doibase 10.1007/JHEP02(2015)054} {\bibfield  {journal} {\bibinfo  {journal}
  {JHEP}\ }\textbf {\bibinfo {volume} {02}},\ \bibinfo {pages} {054} (\bibinfo
  {year} {2015})},\ \Eprint {http://arxiv.org/abs/1407.6359} {arXiv:1407.6359
  [hep-th]} \BibitemShut {NoStop}%
\bibitem [{\citenamefont {Hayashi}\ \emph {et~al.}(2018)\citenamefont
  {Hayashi}, \citenamefont {Jefferson}, \citenamefont {Kim}, \citenamefont
  {Ohmori},\ and\ \citenamefont {Vafa}}]{Hayashi:2019fsa}%
  \BibitemOpen
  \bibfield  {author} {\bibinfo {author} {\bibfnamefont {H.}~\bibnamefont
  {Hayashi}}, \bibinfo {author} {\bibfnamefont {P.}~\bibnamefont {Jefferson}},
  \bibinfo {author} {\bibfnamefont {H.-C.}\ \bibnamefont {Kim}}, \bibinfo
  {author} {\bibfnamefont {K.}~\bibnamefont {Ohmori}}, \ and\ \bibinfo {author}
  {\bibfnamefont {C.}~\bibnamefont {Vafa}},\ }\href {\doibase
  10.4310/sdg.2018.v23.n1.a4} {\bibfield  {journal} {\bibinfo  {journal}
  {Surveys Diff. Geom.}\ }\textbf {\bibinfo {volume} {23}},\ \bibinfo {pages}
  {105} (\bibinfo {year} {2018})},\ \Eprint {http://arxiv.org/abs/1905.00116}
  {arXiv:1905.00116 [hep-th]} \BibitemShut {NoStop}%
\bibitem [{\citenamefont {Eckhard}\ \emph {et~al.}(2020)\citenamefont
  {Eckhard}, \citenamefont {Sch{\"a}fer-Nameki},\ and\ \citenamefont
  {Wang}}]{Eckhard:2020jyr}%
  \BibitemOpen
  \bibfield  {author} {\bibinfo {author} {\bibfnamefont {J.}~\bibnamefont
  {Eckhard}}, \bibinfo {author} {\bibfnamefont {S.}~\bibnamefont
  {Sch{\"a}fer-Nameki}}, \ and\ \bibinfo {author} {\bibfnamefont {Y.-N.}\
  \bibnamefont {Wang}},\ }\href {\doibase 10.1007/JHEP07(2020)199} {\bibfield
  {journal} {\bibinfo  {journal} {JHEP}\ }\textbf {\bibinfo {volume} {07}},\
  \bibinfo {pages} {199} (\bibinfo {year} {2020})},\ \Eprint
  {http://arxiv.org/abs/2004.15007} {arXiv:2004.15007 [hep-th]} \BibitemShut
  {NoStop}%
\bibitem [{\citenamefont {Bhardwaj}(2021{\natexlab{a}})}]{Bhardwaj:2020ruf}%
  \BibitemOpen
  \bibfield  {author} {\bibinfo {author} {\bibfnamefont {L.}~\bibnamefont
  {Bhardwaj}},\ }\href {\doibase 10.1007/JHEP09(2021)186} {\bibfield  {journal}
  {\bibinfo  {journal} {JHEP}\ }\textbf {\bibinfo {volume} {09}},\ \bibinfo
  {pages} {186} (\bibinfo {year} {2021}{\natexlab{a}})},\ \Eprint
  {http://arxiv.org/abs/2010.13230} {arXiv:2010.13230 [hep-th]} \BibitemShut
  {NoStop}%
\bibitem [{\citenamefont {Bhardwaj}(2021{\natexlab{b}})}]{Bhardwaj:2020avz}%
  \BibitemOpen
  \bibfield  {author} {\bibinfo {author} {\bibfnamefont {L.}~\bibnamefont
  {Bhardwaj}},\ }\href {\doibase 10.1007/JHEP04(2021)221} {\bibfield  {journal}
  {\bibinfo  {journal} {JHEP}\ }\textbf {\bibinfo {volume} {04}},\ \bibinfo
  {pages} {221} (\bibinfo {year} {2021}{\natexlab{b}})},\ \Eprint
  {http://arxiv.org/abs/2010.13235} {arXiv:2010.13235 [hep-th]} \BibitemShut
  {NoStop}%
\bibitem [{\citenamefont {Apruzzi}\ \emph {et~al.}(2022)\citenamefont
  {Apruzzi}, \citenamefont {Schafer-Nameki}, \citenamefont {Bhardwaj},\ and\
  \citenamefont {Oh}}]{Apruzzi:2021vcu}%
  \BibitemOpen
  \bibfield  {author} {\bibinfo {author} {\bibfnamefont {F.}~\bibnamefont
  {Apruzzi}}, \bibinfo {author} {\bibfnamefont {S.}~\bibnamefont
  {Schafer-Nameki}}, \bibinfo {author} {\bibfnamefont {L.}~\bibnamefont
  {Bhardwaj}}, \ and\ \bibinfo {author} {\bibfnamefont {J.}~\bibnamefont
  {Oh}},\ }\href {\doibase 10.21468/SciPostPhys.13.2.024} {\bibfield  {journal}
  {\bibinfo  {journal} {SciPost Phys.}\ }\textbf {\bibinfo {volume} {13}},\
  \bibinfo {pages} {024} (\bibinfo {year} {2022})},\ \Eprint
  {http://arxiv.org/abs/2105.08724} {arXiv:2105.08724 [hep-th]} \BibitemShut
  {NoStop}%
\bibitem [{\citenamefont {Del~Zotto}\ \emph
  {et~al.}(2022{\natexlab{b}})\citenamefont {Del~Zotto}, \citenamefont {Oh},\
  and\ \citenamefont {Zhou}}]{DelZotto:2021ydd}%
  \BibitemOpen
  \bibfield  {author} {\bibinfo {author} {\bibfnamefont {M.}~\bibnamefont
  {Del~Zotto}}, \bibinfo {author} {\bibfnamefont {J.}~\bibnamefont {Oh}}, \
  and\ \bibinfo {author} {\bibfnamefont {Y.}~\bibnamefont {Zhou}},\ }\href
  {\doibase 10.1007/JHEP08(2022)214} {\bibfield  {journal} {\bibinfo  {journal}
  {JHEP}\ }\textbf {\bibinfo {volume} {08}},\ \bibinfo {pages} {214} (\bibinfo
  {year} {2022}{\natexlab{b}})},\ \Eprint {http://arxiv.org/abs/2109.01110}
  {arXiv:2109.01110 [hep-th]} \BibitemShut {NoStop}%
\bibitem [{\citenamefont {Del~Zotto}\ \emph
  {et~al.}(2022{\natexlab{c}})\citenamefont {Del~Zotto}, \citenamefont
  {Garc{\'\i}a~Etxebarria},\ and\ \citenamefont
  {Schafer-Nameki}}]{DelZotto:2022joo}%
  \BibitemOpen
  \bibfield  {author} {\bibinfo {author} {\bibfnamefont {M.}~\bibnamefont
  {Del~Zotto}}, \bibinfo {author} {\bibfnamefont {I.}~\bibnamefont
  {Garc{\'\i}a~Etxebarria}}, \ and\ \bibinfo {author} {\bibfnamefont
  {S.}~\bibnamefont {Schafer-Nameki}},\ }\href {\doibase
  10.21468/SciPostPhys.13.5.105} {\bibfield  {journal} {\bibinfo  {journal}
  {SciPost Phys.}\ }\textbf {\bibinfo {volume} {13}},\ \bibinfo {pages} {105}
  (\bibinfo {year} {2022}{\natexlab{c}})},\ \Eprint
  {http://arxiv.org/abs/2203.10097} {arXiv:2203.10097 [hep-th]} \BibitemShut
  {NoStop}%
\bibitem [{\citenamefont {Acharya}\ \emph {et~al.}(2024)\citenamefont
  {Acharya}, \citenamefont {Del~Zotto}, \citenamefont {Heckman}, \citenamefont
  {Hubner},\ and\ \citenamefont {Torres}}]{Acharya:2023bth}%
  \BibitemOpen
  \bibfield  {author} {\bibinfo {author} {\bibfnamefont {B.~S.}\ \bibnamefont
  {Acharya}}, \bibinfo {author} {\bibfnamefont {M.}~\bibnamefont {Del~Zotto}},
  \bibinfo {author} {\bibfnamefont {J.~J.}\ \bibnamefont {Heckman}}, \bibinfo
  {author} {\bibfnamefont {M.}~\bibnamefont {Hubner}}, \ and\ \bibinfo {author}
  {\bibfnamefont {E.}~\bibnamefont {Torres}},\ }\href {\doibase
  10.4310/bpam.2024.v1.n1.a5} {\bibfield  {journal} {\bibinfo  {journal}
  {Beijing J. Pure Appl. Math.}\ }\textbf {\bibinfo {volume} {1}},\ \bibinfo
  {pages} {273} (\bibinfo {year} {2024})},\ \Eprint
  {http://arxiv.org/abs/2304.03300} {arXiv:2304.03300 [hep-th]} \BibitemShut
  {NoStop}%
\bibitem [{\citenamefont {De~Marco}\ \emph {et~al.}(2024)\citenamefont
  {De~Marco}, \citenamefont {Del~Zotto}, \citenamefont {Graffeo},\ and\
  \citenamefont {Sangiovanni}}]{DeMarco:2023irn}%
  \BibitemOpen
  \bibfield  {author} {\bibinfo {author} {\bibfnamefont {M.}~\bibnamefont
  {De~Marco}}, \bibinfo {author} {\bibfnamefont {M.}~\bibnamefont {Del~Zotto}},
  \bibinfo {author} {\bibfnamefont {M.}~\bibnamefont {Graffeo}}, \ and\
  \bibinfo {author} {\bibfnamefont {A.}~\bibnamefont {Sangiovanni}},\ }\href
  {\doibase 10.1007/JHEP05(2024)306} {\bibfield  {journal} {\bibinfo  {journal}
  {JHEP}\ }\textbf {\bibinfo {volume} {05}},\ \bibinfo {pages} {306} (\bibinfo
  {year} {2024})},\ \bibinfo {note} {[Erratum: JHEP 08, 067 (2024)]},\ \Eprint
  {http://arxiv.org/abs/2311.04984} {arXiv:2311.04984 [hep-th]} \BibitemShut
  {NoStop}%
\bibitem [{\citenamefont {Garc{\'\i}a~Etxebarria}\ and\ \citenamefont
  {Hosseini}(2025)}]{GarciaEtxebarria:2024fuk}%
  \BibitemOpen
  \bibfield  {author} {\bibinfo {author} {\bibfnamefont {I.}~\bibnamefont
  {Garc{\'\i}a~Etxebarria}}\ and\ \bibinfo {author} {\bibfnamefont {S.~S.}\
  \bibnamefont {Hosseini}},\ }\href {\doibase 10.1007/JHEP12(2024)223}
  {\bibfield  {journal} {\bibinfo  {journal} {JHEP}\ }\textbf {\bibinfo
  {volume} {12}},\ \bibinfo {pages} {223} (\bibinfo {year} {2025})},\ \Eprint
  {http://arxiv.org/abs/2404.16028} {arXiv:2404.16028 [hep-th]} \BibitemShut
  {NoStop}%
\bibitem [{\citenamefont {Gagliano}\ and\ \citenamefont
  {Garc{\'\i}a~Etxebarria}(2024)}]{Gagliano:2024off}%
  \BibitemOpen
  \bibfield  {author} {\bibinfo {author} {\bibfnamefont {F.}~\bibnamefont
  {Gagliano}}\ and\ \bibinfo {author} {\bibfnamefont {I.}~\bibnamefont
  {Garc{\'\i}a~Etxebarria}},\ }\href@noop {} {\  (\bibinfo {year} {2024})},\
  \Eprint {http://arxiv.org/abs/2411.15126} {arXiv:2411.15126 [hep-th]}
  \BibitemShut {NoStop}%
\bibitem [{\citenamefont {Harlow}\ and\ \citenamefont
  {Ooguri}(2021)}]{Harlow:2018tng}%
  \BibitemOpen
  \bibfield  {author} {\bibinfo {author} {\bibfnamefont {D.}~\bibnamefont
  {Harlow}}\ and\ \bibinfo {author} {\bibfnamefont {H.}~\bibnamefont
  {Ooguri}},\ }\href {\doibase 10.1007/s00220-021-04040-y} {\bibfield
  {journal} {\bibinfo  {journal} {Commun. Math. Phys.}\ }\textbf {\bibinfo
  {volume} {383}},\ \bibinfo {pages} {1669} (\bibinfo {year} {2021})},\ \Eprint
  {http://arxiv.org/abs/1810.05338} {arXiv:1810.05338 [hep-th]} \BibitemShut
  {NoStop}%
\bibitem [{\citenamefont {Faulkner}\ and\ \citenamefont
  {Iqbal}(2013)}]{Faulkner:2012gt}%
  \BibitemOpen
  \bibfield  {author} {\bibinfo {author} {\bibfnamefont {T.}~\bibnamefont
  {Faulkner}}\ and\ \bibinfo {author} {\bibfnamefont {N.}~\bibnamefont
  {Iqbal}},\ }\href {\doibase 10.1007/JHEP07(2013)060} {\bibfield  {journal}
  {\bibinfo  {journal} {JHEP}\ }\textbf {\bibinfo {volume} {07}},\ \bibinfo
  {pages} {060} (\bibinfo {year} {2013})},\ \Eprint
  {http://arxiv.org/abs/1207.4208} {arXiv:1207.4208 [hep-th]} \BibitemShut
  {NoStop}%
\bibitem [{\citenamefont {Witten}(1998{\natexlab{a}})}]{Witten:1998qj}%
  \BibitemOpen
  \bibfield  {author} {\bibinfo {author} {\bibfnamefont {E.}~\bibnamefont
  {Witten}},\ }\href {\doibase 10.4310/ATMP.1998.v2.n2.a2} {\bibfield
  {journal} {\bibinfo  {journal} {Adv. Theor. Math. Phys.}\ }\textbf {\bibinfo
  {volume} {2}},\ \bibinfo {pages} {253} (\bibinfo {year}
  {1998}{\natexlab{a}})},\ \Eprint {http://arxiv.org/abs/hep-th/9802150}
  {arXiv:hep-th/9802150} \BibitemShut {NoStop}%
\bibitem [{\citenamefont {Bah}\ \emph {et~al.}(2020)\citenamefont {Bah},
  \citenamefont {Bonetti}, \citenamefont {Minasian},\ and\ \citenamefont
  {Nardoni}}]{Bah:2019rgq}%
  \BibitemOpen
  \bibfield  {author} {\bibinfo {author} {\bibfnamefont {I.}~\bibnamefont
  {Bah}}, \bibinfo {author} {\bibfnamefont {F.}~\bibnamefont {Bonetti}},
  \bibinfo {author} {\bibfnamefont {R.}~\bibnamefont {Minasian}}, \ and\
  \bibinfo {author} {\bibfnamefont {E.}~\bibnamefont {Nardoni}},\ }\href
  {\doibase 10.1007/JHEP01(2020)125} {\bibfield  {journal} {\bibinfo  {journal}
  {JHEP}\ }\textbf {\bibinfo {volume} {01}},\ \bibinfo {pages} {125} (\bibinfo
  {year} {2020})},\ \Eprint {http://arxiv.org/abs/1910.04166} {arXiv:1910.04166
  [hep-th]} \BibitemShut {NoStop}%
\bibitem [{\citenamefont {Bergman}\ \emph {et~al.}(2020)\citenamefont
  {Bergman}, \citenamefont {Tachikawa},\ and\ \citenamefont
  {Zafrir}}]{Bergman:2020ifi}%
  \BibitemOpen
  \bibfield  {author} {\bibinfo {author} {\bibfnamefont {O.}~\bibnamefont
  {Bergman}}, \bibinfo {author} {\bibfnamefont {Y.}~\bibnamefont {Tachikawa}},
  \ and\ \bibinfo {author} {\bibfnamefont {G.}~\bibnamefont {Zafrir}},\ }\href
  {\doibase 10.1007/JHEP07(2020)077} {\bibfield  {journal} {\bibinfo  {journal}
  {JHEP}\ }\textbf {\bibinfo {volume} {07}},\ \bibinfo {pages} {077} (\bibinfo
  {year} {2020})},\ \Eprint {http://arxiv.org/abs/2004.05350} {arXiv:2004.05350
  [hep-th]} \BibitemShut {NoStop}%
\bibitem [{\citenamefont {Bah}\ \emph {et~al.}(2021)\citenamefont {Bah},
  \citenamefont {Bonetti},\ and\ \citenamefont {Minasian}}]{Bah:2020uev}%
  \BibitemOpen
  \bibfield  {author} {\bibinfo {author} {\bibfnamefont {I.}~\bibnamefont
  {Bah}}, \bibinfo {author} {\bibfnamefont {F.}~\bibnamefont {Bonetti}}, \ and\
  \bibinfo {author} {\bibfnamefont {R.}~\bibnamefont {Minasian}},\ }\href
  {\doibase 10.1007/JHEP03(2021)196} {\bibfield  {journal} {\bibinfo  {journal}
  {JHEP}\ }\textbf {\bibinfo {volume} {03}},\ \bibinfo {pages} {196} (\bibinfo
  {year} {2021})},\ \Eprint {http://arxiv.org/abs/2007.15003} {arXiv:2007.15003
  [hep-th]} \BibitemShut {NoStop}%
\bibitem [{\citenamefont {Bergman}\ and\ \citenamefont
  {Hirano}(2022)}]{Bergman:2022otk}%
  \BibitemOpen
  \bibfield  {author} {\bibinfo {author} {\bibfnamefont {O.}~\bibnamefont
  {Bergman}}\ and\ \bibinfo {author} {\bibfnamefont {S.}~\bibnamefont
  {Hirano}},\ }\href {\doibase 10.1007/JHEP11(2022)069} {\bibfield  {journal}
  {\bibinfo  {journal} {JHEP}\ }\textbf {\bibinfo {volume} {11}},\ \bibinfo
  {pages} {069} (\bibinfo {year} {2022})},\ \Eprint
  {http://arxiv.org/abs/2208.09396} {arXiv:2208.09396 [hep-th]} \BibitemShut
  {NoStop}%
\bibitem [{\citenamefont {Witten}(1998{\natexlab{b}})}]{Witten:1998wy}%
  \BibitemOpen
  \bibfield  {author} {\bibinfo {author} {\bibfnamefont {E.}~\bibnamefont
  {Witten}},\ }\href {\doibase 10.1088/1126-6708/1998/12/012} {\bibfield
  {journal} {\bibinfo  {journal} {JHEP}\ }\textbf {\bibinfo {volume} {12}},\
  \bibinfo {pages} {012} (\bibinfo {year} {1998}{\natexlab{b}})},\ \Eprint
  {http://arxiv.org/abs/hep-th/9812012} {arXiv:hep-th/9812012} \BibitemShut
  {NoStop}%
\bibitem [{\citenamefont {Belov}\ and\ \citenamefont
  {Moore}(2004)}]{Belov:2004ht}%
  \BibitemOpen
  \bibfield  {author} {\bibinfo {author} {\bibfnamefont {D.}~\bibnamefont
  {Belov}}\ and\ \bibinfo {author} {\bibfnamefont {G.~W.}\ \bibnamefont
  {Moore}},\ }\href@noop {} {\  (\bibinfo {year} {2004})},\ \Eprint
  {http://arxiv.org/abs/hep-th/0412167} {arXiv:hep-th/0412167} \BibitemShut
  {NoStop}%
\bibitem [{\citenamefont {Berkooz}(1998)}]{Berkooz:1998bx}%
  \BibitemOpen
  \bibfield  {author} {\bibinfo {author} {\bibfnamefont {M.}~\bibnamefont
  {Berkooz}},\ }\href {\doibase 10.1016/S0370-2693(98)00913-7} {\bibfield
  {journal} {\bibinfo  {journal} {Phys. Lett. B}\ }\textbf {\bibinfo {volume}
  {437}},\ \bibinfo {pages} {315} (\bibinfo {year} {1998})},\ \Eprint
  {http://arxiv.org/abs/hep-th/9802195} {arXiv:hep-th/9802195} \BibitemShut
  {NoStop}%
\bibitem [{\citenamefont {Polchinski}(2004)}]{Polchinski:2003bq}%
  \BibitemOpen
  \bibfield  {author} {\bibinfo {author} {\bibfnamefont {J.}~\bibnamefont
  {Polchinski}},\ }\href {\doibase 10.1142/S0217751X0401866X} {\bibfield
  {journal} {\bibinfo  {journal} {Int. J. Mod. Phys. A}\ }\textbf {\bibinfo
  {volume} {19S1}},\ \bibinfo {pages} {145} (\bibinfo {year} {2004})},\ \Eprint
  {http://arxiv.org/abs/hep-th/0304042} {arXiv:hep-th/0304042} \BibitemShut
  {NoStop}%
\bibitem [{\citenamefont {Banks}\ and\ \citenamefont
  {Seiberg}(2011)}]{Banks:2010zn}%
  \BibitemOpen
  \bibfield  {author} {\bibinfo {author} {\bibfnamefont {T.}~\bibnamefont
  {Banks}}\ and\ \bibinfo {author} {\bibfnamefont {N.}~\bibnamefont
  {Seiberg}},\ }\href {\doibase 10.1103/PhysRevD.83.084019} {\bibfield
  {journal} {\bibinfo  {journal} {Phys. Rev. D}\ }\textbf {\bibinfo {volume}
  {83}},\ \bibinfo {pages} {084019} (\bibinfo {year} {2011})},\ \Eprint
  {http://arxiv.org/abs/1011.5120} {arXiv:1011.5120 [hep-th]} \BibitemShut
  {NoStop}%
\bibitem [{\citenamefont {Gukov}\ and\ \citenamefont
  {Witten}(2006)}]{Gukov:2006jk}%
  \BibitemOpen
  \bibfield  {author} {\bibinfo {author} {\bibfnamefont {S.}~\bibnamefont
  {Gukov}}\ and\ \bibinfo {author} {\bibfnamefont {E.}~\bibnamefont {Witten}},\
  }\href@noop {} {\  (\bibinfo {year} {2006})},\ \Eprint
  {http://arxiv.org/abs/hep-th/0612073} {arXiv:hep-th/0612073} \BibitemShut
  {NoStop}%
\bibitem [{\citenamefont {Gukov}\ and\ \citenamefont
  {Witten}(2010)}]{Gukov:2008sn}%
  \BibitemOpen
  \bibfield  {author} {\bibinfo {author} {\bibfnamefont {S.}~\bibnamefont
  {Gukov}}\ and\ \bibinfo {author} {\bibfnamefont {E.}~\bibnamefont {Witten}},\
  }\href {\doibase 10.4310/ATMP.2010.v14.n1.a3} {\bibfield  {journal} {\bibinfo
   {journal} {Adv. Theor. Math. Phys.}\ }\textbf {\bibinfo {volume} {14}},\
  \bibinfo {pages} {87} (\bibinfo {year} {2010})},\ \Eprint
  {http://arxiv.org/abs/0804.1561} {arXiv:0804.1561 [hep-th]} \BibitemShut
  {NoStop}%
\bibitem [{\citenamefont {Cattaneo}(1996)}]{Cattaneo:1996pz}%
  \BibitemOpen
  \bibfield  {author} {\bibinfo {author} {\bibfnamefont {A.~S.}\ \bibnamefont
  {Cattaneo}},\ }\href {\doibase 10.1063/1.531595} {\bibfield  {journal}
  {\bibinfo  {journal} {J. Math. Phys.}\ }\textbf {\bibinfo {volume} {37}},\
  \bibinfo {pages} {3684} (\bibinfo {year} {1996})},\ \Eprint
  {http://arxiv.org/abs/q-alg/9602015} {arXiv:q-alg/9602015} \BibitemShut
  {NoStop}%
\bibitem [{\citenamefont {Cattaneo}\ and\ \citenamefont
  {Rossi}(2001)}]{Cattaneo:2000mc}%
  \BibitemOpen
  \bibfield  {author} {\bibinfo {author} {\bibfnamefont {A.~S.}\ \bibnamefont
  {Cattaneo}}\ and\ \bibinfo {author} {\bibfnamefont {C.~A.}\ \bibnamefont
  {Rossi}},\ }\href {\doibase 10.1007/s002200100484} {\bibfield  {journal}
  {\bibinfo  {journal} {Commun. Math. Phys.}\ }\textbf {\bibinfo {volume}
  {221}},\ \bibinfo {pages} {591} (\bibinfo {year} {2001})},\ \Eprint
  {http://arxiv.org/abs/math/0010172} {arXiv:math/0010172} \BibitemShut
  {NoStop}%
\bibitem [{\citenamefont {Cattaneo}\ and\ \citenamefont
  {Rossi}(2005)}]{Cattaneo:2002tk}%
  \BibitemOpen
  \bibfield  {author} {\bibinfo {author} {\bibfnamefont {A.~S.}\ \bibnamefont
  {Cattaneo}}\ and\ \bibinfo {author} {\bibfnamefont {C.~A.}\ \bibnamefont
  {Rossi}},\ }\href {\doibase 10.1007/s00220-005-1339-0} {\bibfield  {journal}
  {\bibinfo  {journal} {Commun. Math. Phys.}\ }\textbf {\bibinfo {volume}
  {256}},\ \bibinfo {pages} {513} (\bibinfo {year} {2005})},\ \Eprint
  {http://arxiv.org/abs/math-ph/0210037} {arXiv:math-ph/0210037} \BibitemShut
  {NoStop}%
\bibitem [{\citenamefont {Bonetti}\ \emph {et~al.}(2025)\citenamefont
  {Bonetti}, \citenamefont {Del~Zotto},\ and\ \citenamefont
  {Minasian}}]{Bonetti:2025dvm}%
  \BibitemOpen
  \bibfield  {author} {\bibinfo {author} {\bibfnamefont {F.}~\bibnamefont
  {Bonetti}}, \bibinfo {author} {\bibfnamefont {M.}~\bibnamefont {Del~Zotto}},
  \ and\ \bibinfo {author} {\bibfnamefont {R.}~\bibnamefont {Minasian}},\
  }\href@noop {} {\  (\bibinfo {year} {2025})},\ \Eprint
  {http://arxiv.org/abs/2509.10343} {arXiv:2509.10343 [hep-th]} \BibitemShut
  {NoStop}%
\bibitem [{\citenamefont {Argurio}\ \emph {et~al.}(2024)\citenamefont
  {Argurio}, \citenamefont {Benini}, \citenamefont {Bertolini}, \citenamefont
  {Galati},\ and\ \citenamefont {Niro}}]{Argurio:2024oym}%
  \BibitemOpen
  \bibfield  {author} {\bibinfo {author} {\bibfnamefont {R.}~\bibnamefont
  {Argurio}}, \bibinfo {author} {\bibfnamefont {F.}~\bibnamefont {Benini}},
  \bibinfo {author} {\bibfnamefont {M.}~\bibnamefont {Bertolini}}, \bibinfo
  {author} {\bibfnamefont {G.}~\bibnamefont {Galati}}, \ and\ \bibinfo {author}
  {\bibfnamefont {P.}~\bibnamefont {Niro}},\ }\href {\doibase
  10.1007/JHEP07(2024)130} {\bibfield  {journal} {\bibinfo  {journal} {JHEP}\
  }\textbf {\bibinfo {volume} {07}},\ \bibinfo {pages} {130} (\bibinfo {year}
  {2024})},\ \Eprint {http://arxiv.org/abs/2404.06601} {arXiv:2404.06601
  [hep-th]} \BibitemShut {NoStop}%
\bibitem [{\citenamefont {Lin}\ \emph {et~al.}(2023)\citenamefont {Lin},
  \citenamefont {Okada}, \citenamefont {Seifnashri},\ and\ \citenamefont
  {Tachikawa}}]{Lin:2022dhv}%
  \BibitemOpen
  \bibfield  {author} {\bibinfo {author} {\bibfnamefont {Y.-H.}\ \bibnamefont
  {Lin}}, \bibinfo {author} {\bibfnamefont {M.}~\bibnamefont {Okada}}, \bibinfo
  {author} {\bibfnamefont {S.}~\bibnamefont {Seifnashri}}, \ and\ \bibinfo
  {author} {\bibfnamefont {Y.}~\bibnamefont {Tachikawa}},\ }\href {\doibase
  10.1007/JHEP03(2023)094} {\bibfield  {journal} {\bibinfo  {journal} {JHEP}\
  }\textbf {\bibinfo {volume} {03}},\ \bibinfo {pages} {094} (\bibinfo {year}
  {2023})},\ \Eprint {http://arxiv.org/abs/2208.05495} {arXiv:2208.05495
  [hep-th]} \BibitemShut {NoStop}%
\bibitem [{\citenamefont {Bartsch}\ \emph
  {et~al.}(2023{\natexlab{b}})\citenamefont {Bartsch}, \citenamefont
  {Bullimore},\ and\ \citenamefont {Grigoletto}}]{Bartsch:2023pzl}%
  \BibitemOpen
  \bibfield  {author} {\bibinfo {author} {\bibfnamefont {T.}~\bibnamefont
  {Bartsch}}, \bibinfo {author} {\bibfnamefont {M.}~\bibnamefont {Bullimore}},
  \ and\ \bibinfo {author} {\bibfnamefont {A.}~\bibnamefont {Grigoletto}},\
  }\href@noop {} {\  (\bibinfo {year} {2023}{\natexlab{b}})},\ \Eprint
  {http://arxiv.org/abs/2304.03789} {arXiv:2304.03789 [hep-th]} \BibitemShut
  {NoStop}%
\bibitem [{\citenamefont {Bhardwaj}\ and\ \citenamefont
  {Schafer-Nameki}(2024)}]{Bhardwaj:2023wzd}%
  \BibitemOpen
  \bibfield  {author} {\bibinfo {author} {\bibfnamefont {L.}~\bibnamefont
  {Bhardwaj}}\ and\ \bibinfo {author} {\bibfnamefont {S.}~\bibnamefont
  {Schafer-Nameki}},\ }\href {\doibase 10.21468/SciPostPhys.16.4.093}
  {\bibfield  {journal} {\bibinfo  {journal} {SciPost Phys.}\ }\textbf
  {\bibinfo {volume} {16}},\ \bibinfo {pages} {093} (\bibinfo {year} {2024})},\
  \Eprint {http://arxiv.org/abs/2304.02660} {arXiv:2304.02660 [hep-th]}
  \BibitemShut {NoStop}%
\bibitem [{\citenamefont {Braun}\ \emph {et~al.}(2024)\citenamefont {Braun},
  \citenamefont {Sabag}, \citenamefont {Sacchi},\ and\ \citenamefont
  {Schafer-Nameki}}]{Braun:2023fqa}%
  \BibitemOpen
  \bibfield  {author} {\bibinfo {author} {\bibfnamefont {A.~P.}\ \bibnamefont
  {Braun}}, \bibinfo {author} {\bibfnamefont {E.}~\bibnamefont {Sabag}},
  \bibinfo {author} {\bibfnamefont {M.}~\bibnamefont {Sacchi}}, \ and\ \bibinfo
  {author} {\bibfnamefont {S.}~\bibnamefont {Schafer-Nameki}},\ }\href
  {\doibase 10.21468/SciPostPhys.17.4.102} {\bibfield  {journal} {\bibinfo
  {journal} {SciPost Phys.}\ }\textbf {\bibinfo {volume} {17}},\ \bibinfo
  {pages} {102} (\bibinfo {year} {2024})},\ \Eprint
  {http://arxiv.org/abs/2304.01193} {arXiv:2304.01193 [hep-th]} \BibitemShut
  {NoStop}%
\bibitem [{\citenamefont {'t~Hooft}(1981)}]{tHooft:1981bkw}%
  \BibitemOpen
  \bibfield  {author} {\bibinfo {author} {\bibfnamefont {G.}~\bibnamefont
  {'t~Hooft}},\ }\href {\doibase 10.1016/0550-3213(81)90442-9} {\bibfield
  {journal} {\bibinfo  {journal} {Nucl. Phys. B}\ }\textbf {\bibinfo {volume}
  {190}},\ \bibinfo {pages} {455} (\bibinfo {year} {1981})}\BibitemShut
  {NoStop}%
\bibitem [{\citenamefont {Copetti}\ \emph {et~al.}(2025)\citenamefont
  {Copetti}, \citenamefont {Del~Zotto}, \citenamefont {Ohmori},\ and\
  \citenamefont {Wang}}]{Copetti:2023mcq}%
  \BibitemOpen
  \bibfield  {author} {\bibinfo {author} {\bibfnamefont {C.}~\bibnamefont
  {Copetti}}, \bibinfo {author} {\bibfnamefont {M.}~\bibnamefont {Del~Zotto}},
  \bibinfo {author} {\bibfnamefont {K.}~\bibnamefont {Ohmori}}, \ and\ \bibinfo
  {author} {\bibfnamefont {Y.}~\bibnamefont {Wang}},\ }\href {\doibase
  10.1007/s00220-024-05227-9} {\bibfield  {journal} {\bibinfo  {journal}
  {Commun. Math. Phys.}\ }\textbf {\bibinfo {volume} {406}},\ \bibinfo {pages}
  {73} (\bibinfo {year} {2025})},\ \Eprint {http://arxiv.org/abs/2305.18282}
  {arXiv:2305.18282 [hep-th]} \BibitemShut {NoStop}%
\end{thebibliography}%
\bibliographystyle{apsrev4-1}

\end{document}